\documentclass[fleqn,usenatbib]{mnras}

\usepackage{newtxtext,newtxmath}
\usepackage[T1]{fontenc}

\DeclareRobustCommand{\VAN}[3]{#2}
\let\VANthebibliography\thebibliography
\def\thebibliography{\DeclareRobustCommand{\VAN}[3]{##3}\VANthebibliography}

\usepackage{graphicx}	% Including figure files
\usepackage{amsmath}	% Advanced maths commands
\graphicspath{{./}{figures/}}

\newcommand\EriII{Eri {\sc II}}
\newcommand{\tauSFE}{\tau_\text{SFE}}
\newcommand{\tauSFEtext}{$\tau_\text{SFE}$}
\newcommand{\logtauSFE}{\log_{10}\tau_\text{SFE}}
\newcommand{\logtauSFEtext}{$\log_{10}\tau_\text{SFE}$}
\newcommand{\tauSFH}{\tau_\text{SFH}}
\newcommand{\tauSFHtext}{$\tau_\text{SFH}$}
\newcommand{\ttrunc}{t_\text{trunc}}
\newcommand{\ttrunctext}{$t_\text{trunc}$}
\newcommand{\fret}{f_\text{ret}}
\newcommand{\frettext}{$f_\text{ret}$}

\newcommand{\fretIa}{f_\text{ret}^\text{Ia}}

\newcommand{\FeH}{\text{[Fe/H]}}
\newcommand{\FeHtext}{[Fe/H]}
\newcommand{\MgFe}{\text{[Mg/Fe]}}
\newcommand{\alphaFe}{[\alpha/\text{Fe}]}
\newcommand{\alphaFetext}{[$\alpha$/Fe]}
\newcommand{\MgFetext}{[Mg/Fe]}
\newcommand{\yMgCC}{y_{\text{Mg}}^{\text{cc}}}
\newcommand{\yFeCC}{y_{\text{Fe}}^{\text{cc}}}
\newcommand{\yMgIa}{y_{\text{Mg}}^{\text{Ia}}}
\newcommand{\yFeIa}{y_{\text{Fe}}^{\text{Ia}}}

\newcommand{\yFeCCtext}{$y_{\text{Fe}}^{\text{cc}}$}

\newcommand{\yFeIatext}{$y_{\text{Fe}}^{\text{Ia}}$}

\defcitealias{fu:2022}{F22}
\newcommand{\citetFu}{\citetalias{fu:2022}}
\defcitealias{weinberg:2017}{WAF17}
\newcommand{\citetWAF}{\citetalias{weinberg:2017}}
\defcitealias{johnson:2022b}{J22}
\newcommand{\citetJohnson}{\citetalias{johnson:2022b}}
\defcitealias{planck:2020}{Planck Collaboration (2020)}
\newcommand{\citetPlanck}{\citetalias{planck:2020}}

\title[Outflows and Star Formation in \EriII]{Strong Outflows and Inefficient Star Formation in the Reionization-era \\ Ultra-faint Dwarf Galaxy Eridanus {\sc II}}

\author[N. R. Sandford et al.]{
Nathan R. Sandford,$^{1}$\thanks{E-mail: nathan\_sandford@berkeley.edu}
David H. Weinberg,$^{2}$
Daniel R. Weisz,$^{1}$
and Sal Wanying Fu$^{1}$
\\
$^{1}$Department of Astronomy, University of California Berkeley, Berkeley, CA 94720, USA\\
$^{2}$Department of Astronomy and Center for Cosmology and AstroParticle Physics, The Ohio State University, Columbus, OH 43210, USA
}

\date{Accepted XXX. Received YYY; in original form ZZZ}

\pubyear{2023}

\begin{document}
\label{firstpage}
\pagerange{\pageref{firstpage}--\pageref{lastpage}}
\maketitle

\begin{abstract}
We present novel constraints on the underlying galaxy formation physics (e.g., mass loading factor, star formation history, metal retention) at $z\gtrsim7$ for the low-mass ($M_*\sim10^5$ M$_\odot$) Local Group ultra-faint dwarf galaxy (UFD) Eridanus {\sc II} (\EriII). Using a hierarchical Bayesian framework, we apply a one-zone chemical evolution model to \EriII's CaHK-based photometric metallicity distribution function (MDF; \FeHtext) and find that the evolution of \EriII\ is well-characterized by a short, exponentially declining star-formation history  ($\tauSFH=0.39\pm_{0.13}^{0.18}$ Gyr), a low star-formation efficiency ($\tauSFE=27.56\pm_{12.92}^{25.14}$ Gyr), and a large mass-loading factor ($\eta=194.53\pm_{42.67}^{33.37}$). Our results are consistent with \EriII\ forming the majority of its stars before the end of reionization. The large mass-loading factor implies strong outflows in the early history of \EriII\ and is in good agreement with theoretical predictions for the mass-scaling of galactic winds.
It also results in the ejection of $>$90\% of the metals produced in \EriII. We make predictions for the distribution of \MgFetext-\FeHtext\ in \EriII\ as well as the prevalence of ultra metal-poor stars, both of which can be tested by future chemical abundance measurements. Spectroscopic follow-up of the highest metallicity stars in \EriII\ ($\FeH > -2$) will greatly improve model constraints. Our new framework can readily be applied to all UFDs throughout the Local Group, providing new insights into the underlying physics governing the evolution of the faintest galaxies in the reionization era.
\end{abstract}


%%%%%%%%%%%%%%%%%%%%%%%%%%%%%%%%%%%%%%%%%%%%%%%%%%

%%%%%%%%%%%%%%%%% BODY OF PAPER %%%%%%%%%%%%%%%%%%

\section{Introduction}
\label{sec:introduction}
At the faintest end of the galaxy luminosity function, ultra-faint dwarf galaxies (UFDs) are some of the oldest ($\gtrsim$13 Gyr), lowest mass ($M_* \lesssim 10^6$ M$_\odot$), most metal-poor ($\text{[Fe/H]} \lesssim -2.0$), and dark matter-dominated ($M/L \gtrsim 100$) systems in the Universe \citep[e.g.,][and references therein]{simon:2019}. As such, Local Group (LG) UFDs and their stellar populations provide a powerful lens through which to study a wide range of astrophysics from the nature of dark matter to star formation, stellar evolution, and chemical enrichment in the early Universe before the epoch of re-ionization. 

Eridanus {\sc II} (\EriII; $M_V=-7.1$), initially discovered in the Dark Energy Survey by \citet{bechtol:2015} and \citet{koposov:2015}, is an ideal UFD to study low-mass galaxy evolution at early times. Its dynamical mass ($M_{1/2}=1.2\pm^{0.4}_{0.3}\times10^{7}$ M$_\odot$) and stellar metallicity distribution function (MDF; $\langle\text{[Fe/H]}\rangle=-2.38\pm0.13$ and $\sigma_\text{[Fe/H]}=0.47\pm^{0.12}_{0.09}$) measured from calcium triplet (CaT) observations of \EriII's brightest red giant branch (RGB) stars confirm its status as a metal-poor dark matter-dominated dwarf galaxy \citep{li:2017}. Later spectroscopic and variable star studies provided independent confirmation of \EriII's metal-poor dark matter-dominated nature \citep{zoutendijk:2020,zoutendijk:2021,martinez-vazquez:2021}. Meanwhile, its star formation history (SFH) measured from deep broadband imaging is consistent with \EriII\ forming nearly all of its stellar mass ($\sim2\times10^5$ M$_\odot$) in a short ($<500$ Myr) burst over 13 Gyr ago, making it a true relic of the pre-reionization era \citep{simon:2021,gallart:2021}. Further, its current distance at $\sim350$ kpc \citep{crnojevic:2016,li:2017,martinez-vazquez:2021, simon:2021} and its orbit inferred from Gaia eDR3 proper motions place \EriII\ at first infall into the Milky Way (MW), indicating that it likely evolved in isolation and thus removing the need to account for ram pressure stripping or tidal interactions during its evolution \citep{battaglia:2022, fu:2022}.

Recently, \citet[][hereafter \citetFu{}]{fu:2022} presented newly measured [Fe/H] abundances for 60 \EriII\ RGB stars from deep narrowband photometry of the calcium H\&K doublet (CaHK) acquired with the \textit{Hubble Space Telescope} (\textit{HST}). These observations roughly quadrupled the number of \EriII\ stars with known metallicities, substantially improving the sampling of \EriII's MDF measured from the CaT observations of \citet{li:2017}. \citetFu{} found \EriII's MDF to be characterized by a mean metallicity of $\langle\text{[Fe/H]}\rangle=-2.50\pm0.07$ with a dispersion of $\sigma_\text{[Fe/H]}=0.42\pm0.06$. While \citetFu{} fit simple ``closed box" and ``leaky box" chemical evolution models to the \EriII\ MDF, constraints on the physical processes (e.g., star formation and outflows) governing the galaxy's chemical evolution have yet to be attempted.

Here we use the analytic one-zone galactic chemical evolution models first presented in \citet[][hereafter \citetWAF]{weinberg:2017} to fit the MDF of \EriII\ in a hierarchical Bayesian framework that can be applied uniformly to the MDFs of all observed UFDs, present and future. The key assumption of these models is that the star-forming gas reservoir of \EriII\ is efficiently mixed and can therefore be approximated as chemically homogeneous at any given time. The models enable us to infer key galactic evolution parameters for \EriII, including the star formation efficiency (SFE), star formation history (SFH) timescale, and the mass loading factor for \EriII\ and place them in context of past observational and theoretical low-mass galaxy studies.  

The structure of this paper is as follows. In \S\ref{sec:data}, we summarize the data included in our analysis. In Section \ref{sec:methods}, we describe our chemical evolution model and fitting techniques. We present and discuss our results in Sections \ref{sec:results} and \ref{sec:discussion} respectively before concluding in Section \ref{sec:conclusion}. Throughout this work when converting between redshift and age, we assume the flat $\Lambda$CDM cosmology of \citetPlanck{}.

\section{Data}
\label{sec:data}
In this study, we use the iron abundances, \FeHtext, of 60 stars in the UFD galaxy \EriII\ measured by \citetFu{} from \textit{HST} CaHK narrowband photometry (WFC3/UVIS F395) in conjunction with archival broadband photometry (ACS/WFC 475W and F814W). This sample contains only RGB stars with $\text{F475}\lesssim24$ in the inner 260 pc region of \EriII\ and excludes all stars within 2 half-light radii, $r_h$, of the galaxy's singular star cluster.

\citetFu{} fit the CaHK color index\footnote{Defined as $\text{F395N} - \text{F475W} - 1.5(\text{F475W}-\text{F814W})$} of each star using 13 Gyr old mono-metallic $\alpha$-enhanced MIST isochrones \citep{choi:2016, dotter:2016} to infer \FeHtext\ for each star in their sample. Employing Bayesian techniques enabled them to recover the posterior distribution of \FeHtext\ for each star, assuming a flat prior. Many stars in their sample exhibit non-Gaussian uncertainties in \FeHtext\ with long tails towards low metallicity, which occur as a result of less distinguishable CaHK absorption features in metal-poor stars. A few stars have \FeHtext\ posteriors that truncate at $\FeH=-4.0$ due to the limited extent of the MIST model grid. To capture this non-Gaussianity in our analysis, we approximate the sampled posterior distribution of each star from the MCMC chains of \citetFu{} using bounded Gaussian kernel density estimation (KDE).

\begin{figure*} 
    \begin{center}
    \includegraphics[width=\textwidth]{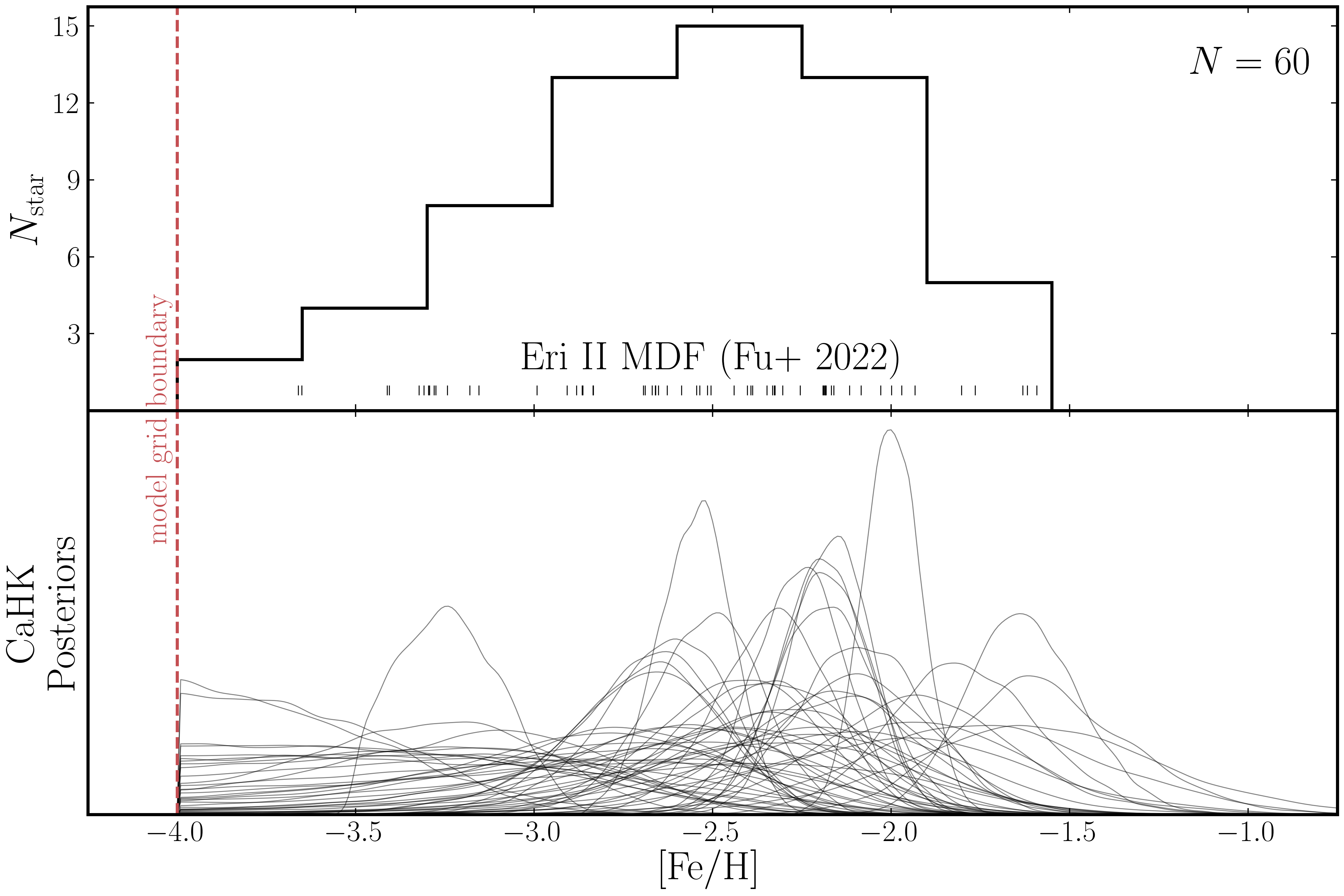}
    \caption{
        Top: Distribution of narrowband CaHK metallicity measurements for 60 RGB stars in the Eri II UFD reported by F22. The median measurement uncertainty (0.35 dex) is reflected in the choice of bin size. Each individual measurement is represented by a tick in the rug plot. 
        Bottom: Posterior [Fe/H] distributions for each star approximated by applying bounded Gaussian kernel density estimation to the MCMC samples of F22. The posteriors of several stars exhibit long tails towards low metallicity and/or truncation at the limit of the MIST model grid ($\text{[Fe/H]}=-4.0$).
        \label{fig:EriII_CaHK_MDF}
    }
    \end{center}
\end{figure*}

Figure \ref{fig:EriII_CaHK_MDF} shows the MDF of the 60 \EriII\ RGB stars in our sample. In the top panel, we plot the MDF as a histogram using the posterior median \FeHtext\ values for each star reported by \citetFu{}. A bin width of 0.35 dex is chosen to match the median measurement uncertainty. In the bottom panel, we display the approximated CaHK \FeHtext\ posterior distribution for each star. These will later be used as priors in our analysis (see Section \ref{sec:likelihood_priors}).

\section{Methods}
\label{sec:methods}

\subsection{Chemical Evolution Model}
\label{sec:model}
We adopt the galactic chemical evolution framework presented and discussed extensively in \citetWAF{}. In brief, this analytic model tracks the time evolution of abundances in a fully mixed (one-zone) system experiencing gas accretion, star formation, supernova enrichment, and outflows. Relative to previous analytic models, the key innovation of the \citetWAF{} model is its ability to separately track both rapid enrichment from core collapse supernovae (CC SNe) and delayed enrichment from Type Ia supernovae (SNe Ia). In the limit of prompt enrichment and no gas accretion, the model approaches ``closed box'' (no outflow) or ``leaky box'' scenarios, but the behavior in this limit is quite different from that of models with ongoing accretion. A complete description of the model and its input parameters can be found in \citetWAF{} (see especially their Table 1). We summarize parameter choices for our fiducial \EriII\ model below and in Table \ref{tab:model_par}. We also consider several alternative models with variations on the fiducial choices, which we describe in Section \ref{sec:alt_models}. A discussion of key model assumptions and their potential impact on the interpretation of \EriII's chemical evolution is presented in Section \ref{sec:limitations}.

\subsubsection{Star Formation}
Motivated by the star formation history (SFH) measured for \EriII\ by \citet{simon:2021} and \citet{gallart:2021}, we adopt a truncated exponentially declining star formation rate
\begin{equation}\label{eq:SFH}
    \dot{M}_*\propto
    \begin{cases}
        \exp{(-t/\tauSFH)},& \text{if } t \leq \ttrunc \\
        0,& \text{if } t > \ttrunc
    \end{cases}
\end{equation}
where \tauSFHtext\ is the SFH timescale and \ttrunctext\ is the time at which all star formation ceases. The sharp truncation of the SFH is adopted to simulate the abrupt quenching of low-mass galaxies (e.g., from ram pressure stripping or reionization).
We leave both \tauSFHtext\ and \ttrunctext\ as free parameters.

The conversion of gas into stars is governed by a linear star formation law characterized by the star formation efficiency (SFE) timescale (or inverse SFE) according to
\begin{equation}
    \tauSFE\equiv \text{SFE}^{-1}\equiv M_g/\dot{M}_*,
\end{equation}
where $M_g$ and $\dot{M}_*$ are the gas mass and star formation rate (SFR) respectively. We leave \tauSFEtext\ as a free parameter. (In \citetWAF{}, the SFE timescale is denoted $\tau_*$.)

\subsubsection{Gas Flows}
The mass recycling fraction, $r$, sets the fraction of mass formed into stars that is immediately returned to the ISM without further chemical enrichment by CCSNe and asymptotic giant branch (AGB) stars. Because this recycling is not a source of new metals, its main effect is to slow the rate at which metals in the ISM are depleted by star formation. We adopt a recycling fraction $r=0.37$, which is appropriate for a \citet{kroupa:2001} IMF after 1 Gyr. As shown by \citetWAF{}, treating this recycling as instantaneous is an accurate approximation, because much of the recycled material originates from stars with short lifetimes (see their Fig.\ 7). Moreover, the effect of this approximation is small when the metallicity is low or when galactic winds are important as is the case for \EriII.

Gas ejected from the ISM by stellar feedback (i.e., CCSNe and AGB winds) scales linearly with the SFR according to
\begin{equation}
    \eta = \dot{M}_\text{outflow}/\dot{M}_*,
\end{equation}
where $\eta$ is the mass-loading factor. We leave $\eta$ as a free parameter.

Gas inflow is specified implicitly in the model through the provided SFH, SFE, mass recycling fraction, and mass-loading factor such that the depletion of gas by star formation and outflows is sufficiently balanced to maintain the SFR given in Equation \ref{eq:SFH}. \citetWAF{} demonstrates that the gas infall rate can be obtained analytically in terms of other model parameters as
\begin{equation} \label{eq:mdotinfall}
    \dot{M}_\text{inf}=(1+\eta-r)\dot{M}_* + \tauSFE \ddot{M}_*
\end{equation}
(see their Equation 9). For our exponential SFH, $\ddot{M}_* = -\dot{M}_*/\tauSFH$. We assume accreted gas is pristine and free of previous enrichment.

\subsubsection{Chemical Enrichment}
Enrichment from CC SNe is assumed to occur instantaneously following star formation. 
Enrichment from SNe Ia, on the other hand, is assumed to follow a delay time distribution (DTD) that accurately approximates the $t^{-1.1}$ power-law found empirically by \citet{maoz:2012}\footnote{As discussed in \citetWAF{}, we approximate the power-law distribution using a sum of two exponentials to allow for an analytic solution.}.
%Enrichment from SNe Ia is assumed to follow a sum of exponentials (allowing analytic solution) that accurately approximates a $t^{-1.1}$ power-law delay time distribution (DTD) as favored in \citet{maoz:2012}. 
We adopt a minimum time delay, $t_D$, of 0.05 Gyr corresponding to the lifetime of the most massive white dwarf progenitors.

The \citetWAF{} model parametrizes chemical enrichment using dimensionless IMF-weighted yield parameters, which are presumed to be independent of metallicity. These yield parameters represent the mass of elements produced per unit mass of star formation. We adopt lower yield values than \citetWAF{}, motivated by the recent study of \citet{rodriguez:2022},  who infer a population-averaged mean Fe yield of 0.058 M$_\odot$ per CC SN. A \citet{kroupa:2001} IMF predicts approximately one $M>8$ M$_\odot$ star per 100 M$_\odot$ of star formation, so this estimate suggests a dimensionless CC SN Fe yield $\yFeCC\approx6\times10^{-4}$, which we adopt for our models. Although our data do not include Mg abundances, we present predictions of \MgFetext\ vs.\ \FeHtext\ that could be tested with future data. We choose $\yMgCC=0.001$, which puts the low-metallicity $\alpha$ ``plateau" at $\MgFe\approx0.5$, roughly consistent with measurements in the MW disk from the H3 Survey  \citep{conroy:2022}. We assume that Mg has no SN Ia contribution, i.e., $\yMgIa=0$. Finally, we choose $\yFeIa=0.0012$ so that models evolved with ``Milky Way disk" parameters similar to \citetWAF{} reach $\MgFe\approx0$ at late times. For an Fe yield of 0.7 M$_\odot$ per SN Ia, this \yFeIatext\ corresponds to $1.7\times10^{-3}$ SNe Ia per M$_\odot$ of star formation, approximately consistent with the rate found by \citet{maoz:2017}. Moderate changes to the yields would change our best fit parameter values, especially for $\eta$, but they would not change our qualitative conclusions. 

The products of CC SNe and SNe Ia that are deposited into the ISM are assumed to mix completely and instantaneously such that they are available for star formation immediately. This simplification, known as the instantaneous mixing approximation, has been shown to be a reasonable assumptions for CC SNe and SNe Ia products in low-mass, ancient galaxies like \EriII\ \citep[e.g.,][]{escala:2018}.

The \citetWAF{} model assumes that outflows are comprised of gas at the ISM metallicity, so that the associated metal loss rate is $\eta\dot{M}_{*}Z_\text{ISM}$. We also consider an alternative formulation in which a fraction of supernova-produced metals are directly ejected from the galaxy and only a fraction \frettext\ are retained within the star-forming ISM. In this case, all yields are multiplied by the factor \frettext, which we assume to be the same for CC SNe and SNe Ia because without \alphaFetext\ measurements we have little leverage to separate the two retention factors. The outflows described by $\eta$ are still assumed to be at the ISM metallicity, but the total metal loss rate is larger because of the direct ejection, which implicitly occurs at a rate $y(1-\fret)\dot{M}_{*}$ for each channel. In our Fiducial model, we fix $\fret=1$, reproducing the scenario in which all supernova-produced metals are deposited initially into the star-forming ISM.

\subsubsection{Initial and Final Conditions}
\label{sec:initial_final_conditions}
Initial conditions of the model are largely set by the aforementioned model parameters. An exponential SFH as assumed in our Fiducial model requires that \EriII\ begin with a non-zero gas mass at $t=0$ Gyr such that
\begin{equation}
    M_{g}(t=0) = \tauSFE \dot{M}_{*}(t=0).
\end{equation}
This initial gas mass is assumed to be primordial in composition (e.g., $Z=0$). The stellar mass of \EriII\ at $t=0$ Gyr is assumed to be zero.

The evolution of the model effectively ends when the SFR is abruptly truncated at $t=\ttrunc$. Within the framework of this model, such a truncation could be achieved by removing all gas from the ISM and shutting off gas accretion (e.g., setting $M_g=\dot{M}_\text{inf} = 0$), by heating gas in the ISM such that it cannot form stars (e.g., setting $\tauSFE=\infty$), or some combination of these effects. Both ram pressure stripping and reionization provide plausible physical explanations for the truncation of star formation, though the latter seems more likely given \EriII's relative isolation. We do not attempt to include a more detailed prescription for star formation truncation as our dataset is of insufficient size and quality to yield meaningful insight.

Finally, all model parameters listed in Table \ref{tab:model_par} are assumed to be constant throughout \EriII's evolution, though we do not expect this to be strictly true in reality. For example, SN yields might vary with stellar metallicity, and the mass-loading factor could decrease over cosmic time as the mass of \EriII's dark matter halo grows. We leave more detailed analysis using time- and metallicity-dependent parameters for future study, noting that the parameters used in this work can be thought of as time-averaged quantities characteristic of \EriII's evolution.

\subsubsection{Constructing the MDF}
The number of stars born as a function of metallicity predicted by the model can be defined using the chain rule in terms of the SFR and the rate of change in \FeHtext\ with time:
\begin{equation}
    \frac{dN}{d\FeH} = \frac{dN/dt}{d\FeH/dt} \propto \frac{\dot{M}_*}{d\FeH/dt}.
    \label{eq:chain}
\end{equation}
We caution that Equation \ref{eq:chain} only holds for a monotonically increasing \FeHtext, which is universally true for the \citetWAF{} model. Time steps of $dt=10^{-5}$ Gyr and metallicity sampling of $d\FeH=0.01$ dex sufficiently minimize numerical artefacts.
To convert between mass fractions of Fe and Mg predicted by the model and solar-scaled abundances [Fe/H] and [Mg/H], we adopt the photospheric abundance scale from \citet{asplund:2009}, corresponding to solar mass fractions of 0.0013 and 0.0056 respectively. This choice of solar abundance scale purposefully matches the solar abundance scale used by the MIST isochrones that underpin the \citetFu{}\ CaHK measurements.

\begin{table*}
	\centering
        \begin{tabular}{llccc}
        \hline \hline
	    Parameter & Description & Value/Priors & Units & References \\
        \hline
	    \multicolumn{5}{c}{Fixed Parameters} \\
	    \hline
	    $\Delta t$            & Time step                                         & $10^{-5}$ & Gyr & ... \\
	    $Z_{\text{Fe},\odot}$ & Solar iron abundance by mass                      & $0.0013$ & ... & [1] \\
	    $Z_{\text{Mg},\odot}$ & Solar magnesium abundance by mass                 & $0.0007$ & ... & [1] \\
	    $y_{\text{Mg}}^{\text{cc}}$              & IMF-integrated CCSN magnesium yield               & $0.0026$ & ... & [2] \\
	    $y_{\text{Fe}}^{\text{cc}}$              & IMF-integrated CCSN iron yield                    & $0.0012$ & ... & [2] \\
	    $y_{\text{Mg}}^{\text{Ia}}$              & IMF-integrated SN Ia magnesium yield              & $0.0$    & ... & [2] \\
	    $y_{\text{Fe}}^{\text{Ia}}$              & IMF-integrated SN Ia iron yield                   & $0.003$  & ... & [2] \\
	    $r$                   & Mass recycling fraction                           & $0.4$   & ... & [3] \\
	    $\alpha_{\text{Ia}}$  & Slope of SN Ia power-law delay time distribution  & $-1.1$   & ... & [4] \\
	    $t_D$                 & Minimum delay time for SNe Ia                     & $0.05$   & Gyr & [3] \\
	    $f_\text{ret}$               & Fraction of newly produced metals retained by the ISM        & $1.0$ & ... & ... \\
	    \hline
	    \multicolumn{5}{c}{Free Parameters} \\
	    \hline
	    $\tau_\text{SFH}$         & star formation history timescale, $\dot{M}_*\propto e^{-t/\tau_\text{SFH}}$       & $\mathcal{TN}(0.7,0.3, 0.0, \infty)$ & Gyr & [5] \\
	    $\tau_\text{SFE}$         & $=M_{g}/\dot{M}_*$, star formation efficiency timescale                   & $\mathcal{U}(0,10^4)$ & Gyr & ... \\
	    $t_\text{trunc}$         & Time of SFH Truncation                                                    & $\mathcal{TN}(1.0,0.5, 0.0, \infty)$ & Gyr & [5] \\
	    $\eta$            & $=\dot{M}_\text{outflow}/\dot{M}_*$, mass-loading factor                  & $\mathcal{U}(0,10^3)$ & ... & .. \\
            \hline
	\end{tabular}
	\caption{
            Fiducial model parameters adopted in this work. Priors for the free parameters are introduced in Section \ref{sec:likelihood_priors}. The implementation of these parameters is described in detail in WAF17. The SN Ia DTD is a sum of two exponentials that accurately approximates a $t^{-1.1}$ power-law. \textbf{References.} [1] \citet{asplund:2009}, [2] \citet{conroy:2022}, [3] \citet{weinberg:2017}, [4] \citet{maoz:2012}, [5] \citet{gallart:2021}
        }
	\label{tab:model_par}
\end{table*}

\subsection{Likelihood and Priors}
\label{sec:likelihood_priors}
We employ Bayesian hierarchical modelling to fit our chemical evolution model to the MDF of \EriII. As in previous analyses of dwarf galaxy MDFs \citep[e.g.,][]{kirby:2011a}, we normalize the metallicity distribution, $dN/d\FeH$, predicted by our chemical evolution model (Equation \ref{eq:chain}) such that $\int_{-\infty}^{\infty}dN/d\FeH~d\FeH = 1$ and adopt it as a probability distribution function (PDF) for the observed stellar abundances. 

\begin{figure} 
    \begin{center}
    \includegraphics[width=\columnwidth]{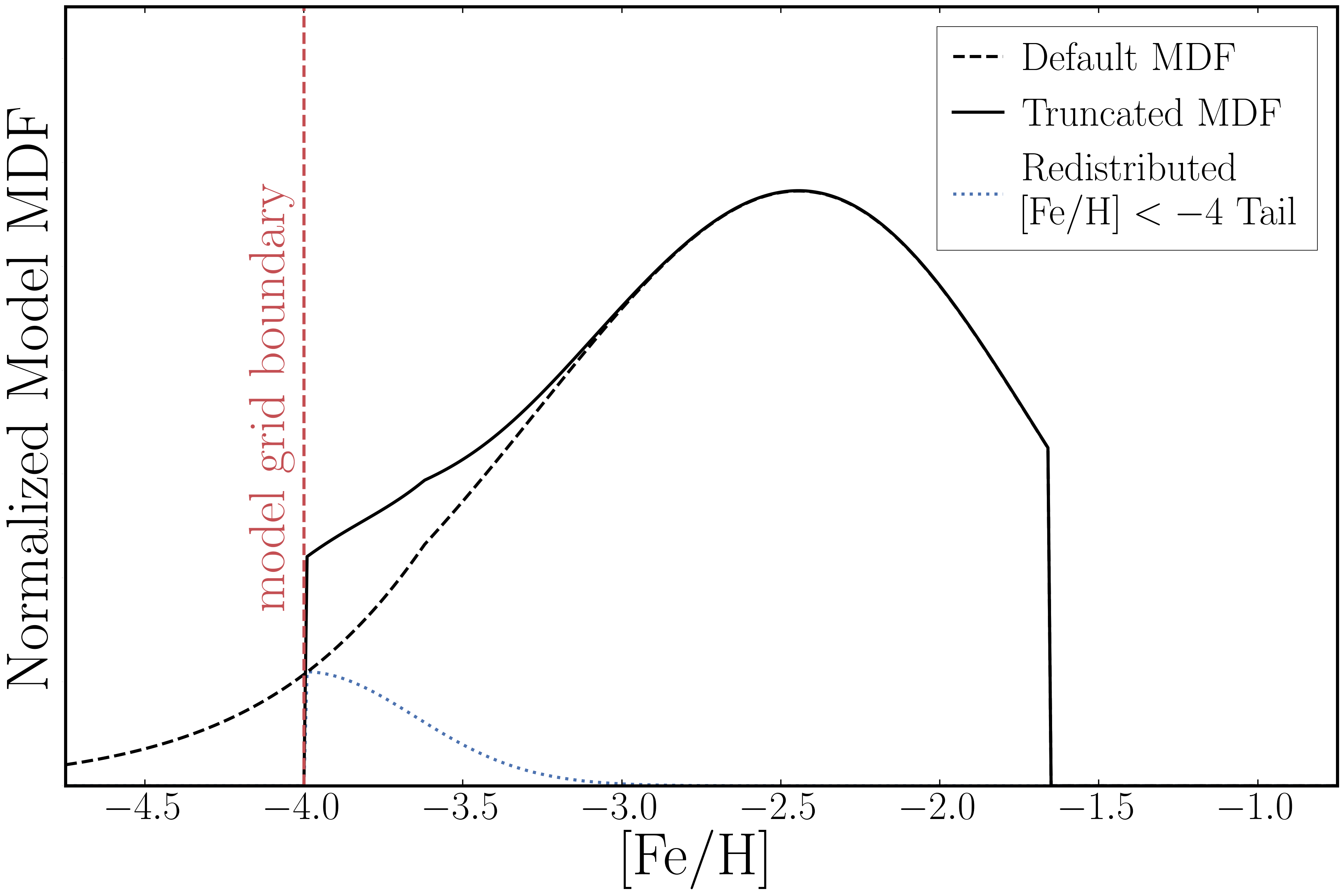}
    \caption{
         Example model MDF (see Section \ref{sec:model}) before and after truncation below $\text{[Fe/H]}<-4$ (dashed and solid black lines respectively). The dotted blue line illustrates the redistribution of the truncated probability following a half-normal distribution with $\sigma=0.35$ dex. This model was generated with the following parameters: $\tau_\text{SFE}=100$ Gyr, $\tau_\text{SFH}=0.5$ Gyr, $t_\text{trunc}=1.0$ Gyr, and $\eta=50$.
        \label{fig:Example_Model_MDF}
    }
    \end{center}
\end{figure}

To account for the lower limit on observed \FeHtext\ imposed by the MIST isochrone grid, we truncate this PDF below $\FeH<-4.0$ and redistribute the truncated mass at the boundary following a half-normal distribution with width $\sigma=0.35$ dex in accordance with the median measurement uncertainty from \citetFu{}. We present an example of a (non-)truncated PDF predicted by the model in Figure \ref{fig:Example_Model_MDF}.

Unlike in previous studies, we do not directly incorporate the observed \FeHtext\ abundances into our likelihood function. Instead, we adopt the posterior distributions from \citetFu{} (described in Section \ref{sec:data}) as priors on the ``latent" \FeHtext\ of each star, which we denote with a prime:
\begin{equation} \label{eq:FeHPrior}
    \text{P}_\text{prior}(\FeH'_i) = \text{P}_\text{Fu22}(\FeH=\FeH'_i | \text{CaHK}_i)
    %\FeH'_i \sim \text{PDF}_{\text{CaHK},i}. \nonumber
\end{equation}
These latent abundances, $\FeH'_i$, are fit simultaneously along with the free model parameters (\tauSFEtext, \tauSFHtext, \ttrunctext, $\eta$, and where relevant \frettext). The total log-likelihood is then
\begin{equation} \label{eq:loglike}
    \ln\mathcal{L} = \sum_{i=1}^{N_*}\ln \frac{d\text{N}}{d\FeH}\biggr|_{\FeH'_i},
\end{equation}
where the sum is over all $N_*$ observed stars and the probability distribution function, $dN/d\FeH$, is evaluated at the latent abundance $\FeH'_i$ of each star. Equation \ref{eq:loglike} ensures that we do not infer an $\FeH'_i$ for any star beyond the maximum \FeHtext\ predicted by the model, while Equation \ref{eq:FeHPrior} penalizes the model for requiring $\FeH'_i$ values in tension with the $\text{CaHK}_i$ measurements.

We adopt a truncated Gaussian prior on the SFH timescale, \tauSFHtext, centered at 0.7 Gyr with width 0.3 Gyr and bounded to be positive definite:
\begin{equation}
    \tauSFH \sim \mathcal{TN}(0.7, 0.3, 0.0, \infty). \nonumber
\end{equation}
This choice of prior is informed by \citet{gallart:2021} who derived a SFH from deep \textit{HST}/ACS photometry of \EriII\ that is peaked at the oldest possible age with half-width at half-maximum (HWHM) of $\sim$0.5 Gyr corresponding to a $\sim$0.7 Gyr e-folding timescale for an exponentially declining SFH. While a negatively skewed prior may be appropriate --\citet{gallart:2021} suggest that the true duration of \EriII's burst of star formation is likely unresolved by their study and could be as short as 100 Myr -- we adopt a Gaussian prior for simplicity.

Similarly, we adopt a positive definite truncated Gaussian prior on the SFH truncation time, \ttrunctext, centered at 1 Gyr with width 0.5 Gyr. 
\begin{equation}
    \ttrunc \sim \mathcal{TN}(1.0, 0.5, 0.0, \infty). \nonumber
\end{equation}
This too is motivated by the lack of evidence found by \citet{gallart:2021} for star formation in \EriII\ within the last $\sim13$ Gyr years.

We utilize broad uniform priors for the remaining model parameters as follows:
\begin{align}
    \logtauSFE  &\sim \mathcal{U}(0.0, 4.0), \nonumber\\
    \eta        &\sim \mathcal{U}(0.0, 10^{3}) \nonumber
\end{align}
Note that we have reparametrized to fit for \tauSFEtext\ in log-space given the large dynamic range we wish to explore.

Together, the sum of the 64 log-priors (4 for the model parameters and 1 for each of the stars' $\FeH'$) and the log-likelihood presented in Equation \ref{eq:loglike} yield the log-posterior distribution that we wish to sample.

\subsection{Alternative Models}
\label{sec:alt_models}
In addition to the Fiducial model described above, we consider several alternative models to build physical intuition and test specific scenarios. We briefly describe the motivation and adjustments for each below. Any parameters not explicitly referenced are identical to those in the Fiducial model (Table \ref{tab:model_par}).

\paragraph{Linear-Exponential SFR Model}
In the Fiducial model, we assume an exponentially declining SFR, which requires a non-zero gas mass at the onset of star formation. With this model, we test an alternative ``linear-exponential" functional form for the SFR functional form, given by
\begin{equation}
    \dot{M}_*\propto
    \begin{cases}
        t\exp{(-t/\tauSFH)},& \text{if } t \leq \ttrunc \\
        0,& \text{if } t > \ttrunc
    \end{cases}.
\end{equation}
In this model, the need for an initial gas reservoir is avoided as the galaxy begins with no star formation at $t=0$ Gyr. Rather, the SFR increases rapidly from zero to its peak at $t=\tauSFH$ before declining more gradually. As in the fiducial case, we adopt a prior for \tauSFHtext\ motivated by the $\sim$0.5 Gyr SFH HWHM measured by \citet{gallart:2021}. For the above linear-exponential SFH, this corresponds to $\tauSFH\sim0.2$ Gyr, and so we use the following truncated Gaussian prior:
\begin{equation}
    \tauSFH \sim \mathcal{TN}(0.2, 0.1, 0.0, \infty). \nonumber
\end{equation}

\paragraph{Constant SFR Model}
With this model, we test whether a constant SFR could reproduce \EriII's MDF. No \tauSFHtext\ is fit for this model as a constant SFR is equivalent to letting $\tauSFH\rightarrow\infty$.

\paragraph{Metal-Loading Model}
In this model, we enable supernova-produced metals to be directly ejected from the galaxy by allowing a non-zero retention factor. Specifically, we adopt a uniform prior on \frettext:
\begin{align}
    \fret  &\sim \mathcal{U}(0.0,1.0). \nonumber
\end{align}
Due to the large degeneracy between \frettext\ and $\eta$, we find it desirable to impose a tight prior on $\eta$, which we force to be roughly 4 times smaller than preferred in the fiducial case (see Section \ref{sec:results}):
\begin{align}
    \eta  &\sim \mathcal{N}(50, 10). \nonumber
\end{align}
While we do achieve converged Monte Carlo chains with \frettext\ and $\eta$ both free (see Section \ref{sec:sampling}), the degeneracy between the parameters makes the results hard to interpret.

\paragraph{High SFE Model}
With this model, we investigate whether the MDF of \EriII\ can be modelled assuming a short SFE timescale of $\logtauSFE\sim0.4$, which is roughly an order of magnitude smaller than preferred by the Fiducial model (see Section \ref{sec:results}). Such a high SFE might be expected if a large fraction ($\sim$75\%) of \EriII's gas was in the molecular phase. We force this enhanced SFE by implementing a tight prior on \logtauSFEtext\ of 
\begin{align}
    \logtauSFE  &\sim \mathcal{N}(0.4, 0.1). \nonumber
\end{align}

\paragraph{Longer SN Ia Delay Model} 
In the Fiducial model, we implement a minimum time delay for SN Ia of $t_\text{D}=0.05$ Gyr corresponding to the lifetimes of the most massive SN Ia progenitors. However, previous chemical evolution studies \citep[e.g.,][]{schonrich:2009, andrews:2017} have adopted a slightly longer time delay of 0.15 Gyr before the first SN Ia. In this model, we set $t_\text{D}=0.15$ Gyr to test the impact of assuming a more delayed onset of SN Ia.

\paragraph{No SN Ia Model}
To evaluate the importance of SN Ia enrichment on the shape of \EriII's MDF, we consider a scenario in which SN Ia do not contribute at all to the enrichment of the galaxy. In this model, we set $\yFeIa=0$ but the same could be accomplished by setting $\fretIa=0$. 

\paragraph{Enhanced SN Ia Model}
In this model we assume the specific SN Ia rate scales with metallicity proportional to $Z^{-0.5}$ as found in the recent analysis of \citet{johnson:2022a}. For \EriII, this scaling would imply an enhancement of the SN Ia rates by roughly an order of magnitude, which we implement by simply increasing the fiducial SN Ia Fe yield \yFeIatext\ by a factor of 10 to $\yFeIa=0.012$.

\subsection{Sampling}
\label{sec:sampling}
To sample our posterior distributions, we employ the \textit{Preconditioned Monte Carlo} (PMC) method for Bayesian inference implemented in the publicly available Python package \texttt{pocoMC}\footnote{\url{https://github.com/minaskar/pocomc}} \citep{karamanis:2022, pocomc:2022}. PMC uses a combination of a normalizing flow with a sequential Monte Carlo sampling scheme to decorrelate and efficiently sample high-dimensional distributions with non-trivial geometry.

We initialize 5000 walkers from the prior distributions described in Section \ref{sec:likelihood_priors}, imposing an arbitrary log-posterior threshold to ensure walkers are not too distant from the bulk of the posterior mass. We adopt default hyperparameters for \texttt{pocoMC}, run until the sampler has converged (i.e., when the ``inverse temperature" $\beta=1$), and then draw an additional 5000 samples for a total of 10,000 samples from the posterior distribution.

\section{Results}
\label{sec:results}

\subsection{Fiducial Fit to \EriII\ MDF}
\label{sec:fiducial_fit}
\begin{figure*} 
    \begin{center}
    \includegraphics[width=\textwidth]{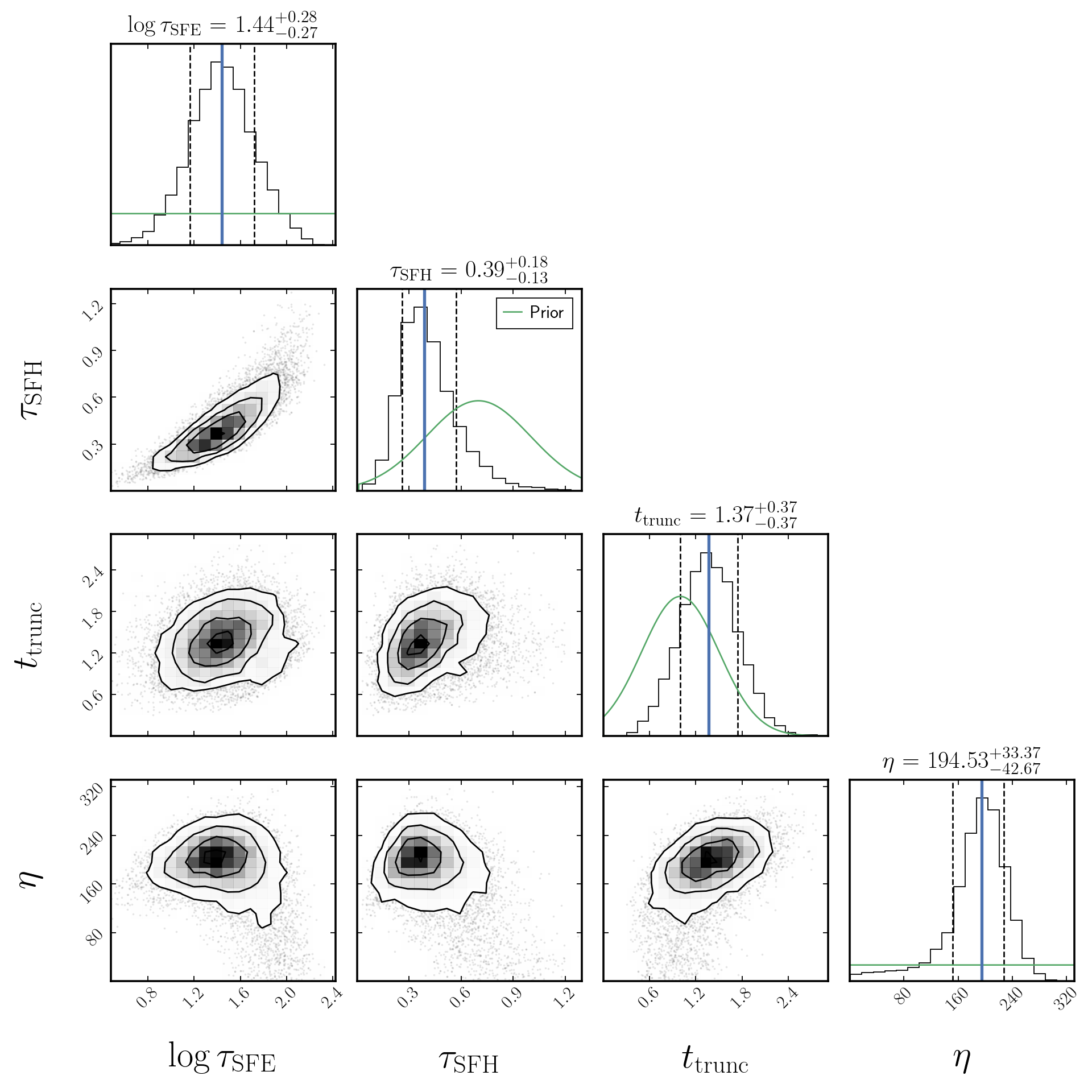}
    \caption{
        Posterior distribution corner plot of the model parameters $\log_{10}\tau_\text{SFE}$, $\tau_\text{SFH}$, $t_\text{trunc}$, and $\eta$. Median values and $1\sigma$ uncertainties from the 16th and 84th percentiles are reported for each label and denoted by solid blue and dashed black lines respectively. The adopted prior distributions are included for reference as solid green lines. The Eri II MDF provides informative constraints on $\log_{10}\tau_\text{SFE}$, $\tau_\text{SFH}$, and $\eta$, while $t_\text{trunc}$\ remains prior-dominated.
        \label{fig:Fiducial_Corner}
    }
    \end{center}
\end{figure*}

We begin by briefly summarizing the recovered posterior distribution for the model parameters \logtauSFEtext, \tauSFHtext, \ttrunctext, and $\eta$, which we display in Figure \ref{fig:Fiducial_Corner}. For each parameter, we report the median of each marginalized posterior distribution (blue lines) and the 16th and 84th percentiles (dashed black lines) in Table \ref{tab:best_fit_par}; for brevity, we refer to the posterior medians as ``best-fit" values hereafter, with the percentile ranges treated as $\pm$1$\sigma$ uncertainties. These are discussed individually in Section \ref{sec:inferred_par}. In short, we find that the MDF of \EriII\ is sufficient to place constraints on \logtauSFEtext\ ($1.44\pm_{0.27}^{0.28}$), \tauSFHtext\ ($0.39\pm_{0.13}^{0.18}$ Gyr), and $\eta$ ($194.53\pm_{42.67}^{33.37}$) but not \ttrunctext\ ($1.37\pm_{0.37}^{0.37}$ Gyr), which remains prior-dominated. We explore the aspects of the MDF's shape that contribute to these constraints (or lack thereof) in Appendix \ref{app:parameter_dependence}.

\begin{figure*} 
    \begin{center}
    \includegraphics[width=\textwidth]{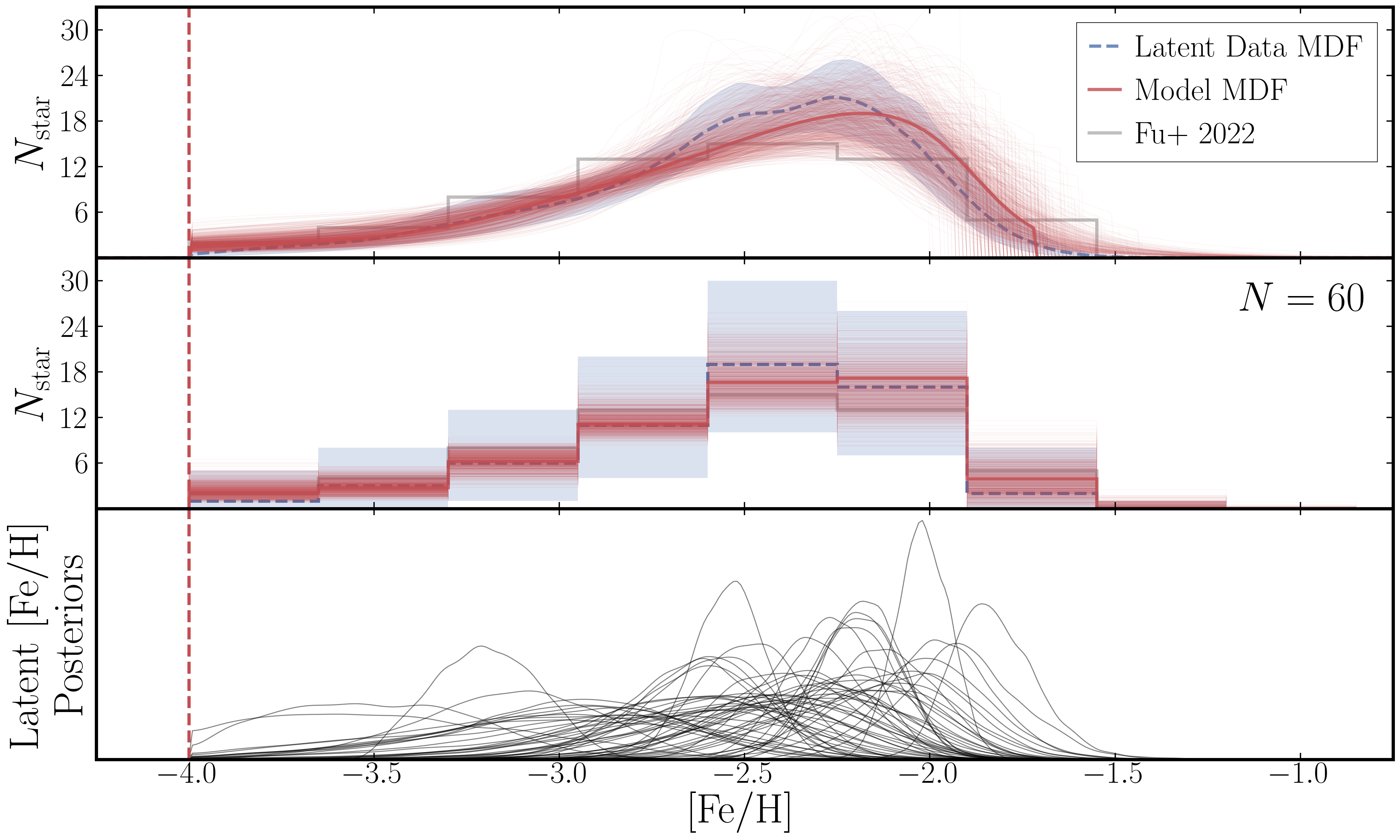}
    \caption{
        Top: Best-fit model MDF (solid red line; computed from the parameter values listed in Table \ref{tab:best_fit_par}) compared to the latent [Fe/H] MDF (dashed blue line). Uncertainties on the latent MDF (blue shaded regions) are estimated via bootstrapping (with replacement) from our sample of stars and drawing each star's latent [Fe/H] posterior distribution. Model MDFs generated from 1000 random draws of the posterior distribution are displayed in thin red lines. We include the observed CaHK [Fe/H] MDF from F22 (solid gray line; same as in Figure \ref{fig:EriII_CaHK_MDF}) for reference but note that the model is not directly fit to this MDF.
        Middle: Same as the top panel but binned for comparison to the F22 CaHK MDF.
        Bottom: Updated posteriors distributions of [Fe/H] for each star in the sample. Compared to the CaHK posteriors presented in the bottom panel of Figure \ref{fig:EriII_CaHK_MDF}, long low-metallicity tails and the degree of truncation at $\text{[Fe/H]}=-4$ are substantially reduced because the model predicts that only a small fraction of stars have such low metallicities. Our fit indicates that it is unlikely that any of the stars in our sample are truly ultra metal-poor stars with $\text{[Fe/H]}\lesssim-4$.
        \label{fig:Fiducial_Model_Fit}
    }
    \end{center}
\end{figure*}

Importantly, we find that our Fiducial model produces realistic predictions for the \EriII\ MDF, which we illustrate in Figure \ref{fig:Fiducial_Model_Fit} using both continuous (top panel) and binned (middle panel) representations. The blue dashed lines in these panels represent the latent MDFs for the best-fit model parameters, which is the sum of the latent posterior distributions of the individual stars (in the top panel) integrated over the bins (in the middle panel). To visualize the uncertainties on the latent MDF, we make a bootstrap selection from our set of 60 stars (allowing replacement) and draw from the star's latent \FeHtext\ posterior distribution, capturing both the uncertainties from finite sample size and the measurement uncertainties for each star. The resulting 95\% confidence interval is depicted by the blue shaded region. The best-fit model MDF (thick red line) is in good agreement with the latent MDF, predicting a negatively skewed distribution with a small low-metallicity tail and little to no truncation below the model grid boundary. We additionally perform a posterior predictive check of our model by generating model MDFs for 1000 random draws from the parameter posteriors (thin red lines), which illustrates the range of MDFs consistent with the uncertainties on our best-fit model parameters. We include the observed CaHK \FeHtext\ MDF from \citetFu{} (solid gray line) for reference, but reiterate that reproducing the latent MDF, not the input CaHK MDF, maximizes the likelihood.

In the bottom panel of Figure \ref{fig:Fiducial_Model_Fit}, we present the posterior distribution for each of the 60 stars' underlying \FeHtext. Compared to the input CaHK posteriors (Figure \ref{fig:EriII_CaHK_MDF}; bottom panel), these updated posteriors exhibit less pronounced low-metallicity tails as well as less frequent and less severe truncation at the model boundary of $\FeH=-4$. The mean metallicity, $\langle\FeH\rangle=-2.52\pm_{0.04}^{0.04}$, and metallicity dispersion, $\sigma_\FeH=0.45\pm_{0.04}^{0.04}$, are still in good agreement with the values found by \citetFu{}.

\begin{table*}
	\centering
        \begin{tabular}{lcccccccc}
	\hline \hline
	    Model & $\log_{10}\tau_\text{SFE}$ & $\tau_\text{SFH}$\ [Gyr] & $t_\text{trunc}$\ [Gyr] & 
            $\eta$ & $f_\text{ret}$ & $t_D$ [Gyr] & $y_{\text{Fe}}^{\text{Ia}}$ & BF \\
	\hline
	    Fiducial               & $1.44\pm_{0.27}^{0.28}$ & $0.39\pm_{0.13}^{0.18}$ & $1.37\pm_{0.37}^{0.37}$ & $194\pm_{33}^{43}$ & $1$ & $0.05$ & $0.0012$ & $1.000$\\
            Linear-Exponential SFR & $1.41\pm_{0.23}^{0.22}$ & $0.21\pm_{0.06}^{0.06}$ & $1.26\pm_{0.32}^{0.36}$ & $186\pm_{40}^{35}$ & $1$ & $0.05$ & $0.0012$ & $3.099$\\
	    Constant SFR           & $1.85\pm_{0.23}^{0.18}$ & $\infty$                & $1.03\pm_{0.34}^{0.37}$ & $17\pm_{12}^{21}$  & $1$ & $0.05$ & $0.0012$ & $0.232$\\
	    Metal-Loading          & $1.00\pm_{0.34}^{0.48}$ & $0.47\pm_{0.16}^{0.19}$ & $1.26\pm_{0.38}^{0.39}$ & $52\pm_{8}^{8}$    & $0.32\pm_{0.07}^{0.22}$ & $0.05$ & $0.0012$ & $0.711$\\
	    High SFE               & $0.47\pm_{0.08}^{0.08}$ & $0.08\pm_{0.08}^{0.04}$ & $1.06\pm_{0.37}^{0.39}$ & $145\pm_{30}^{37}$ & $1$ & $0.05$ & $0.0012$ & $0.002$ \\
	    Longer SN Ia Delay     & $1.44\pm_{0.21}^{0.23}$ & $0.49\pm_{0.14}^{0.16}$ & $1.33\pm_{0.34}^{0.35}$ & $155\pm_{35}^{31}$ & $1$ & $0.15$ & $0.0012$ & $1.018$\\
	    No SN Ia               & $1.57\pm_{0.18}^{0.23}$ & $0.45\pm_{0.18}^{0.21}$ & $1.26\pm_{0.33}^{0.36}$ & $71\pm_{39}^{21}$  & $1$ & $0.05$ & $0.0000$ & $2.239$\\
	    Enhanced SN Ia         & $1.44\pm_{0.31}^{0.48}$ & $0.48\pm_{0.12}^{0.18}$ & $1.06\pm_{0.27}^{0.32}$ & $879\pm_{91}^{61}$ & $1$ & $0.05$ & $0.0120$ & $0.024$ \\
            \hline
        \end{tabular}
	\caption{
	    Inferred Eri II parameters. Median values and $1\sigma$ uncertainties inferred for the model parameters from the fiducial and alternative model fits. Values without uncertainties were held fixed. The estimated Bayes factor relative to the Fiducial model is presented in the right-most column.
	}
	\label{tab:best_fit_par}
\end{table*}

\subsubsection{Inferred Parameters of \EriII}
\label{sec:inferred_par}

\paragraph{Star Formation Efficiency}
We infer the log-SFE timescale of \EriII\ to be $\logtauSFE=1.44\pm_{0.27}^{0.28}$ ($\tauSFE=27.56\pm_{12.92}^{25.14}$ Gyr) or in terms of the SFE ($\tauSFE^{-1}$): $\text{SFE}=0.036\pm_{0.017}^{0.032}$ Gyr$^{-1}$. This timescale is quite large compared to the SFE timescale of molecular gas ($\tauSFE=2$ Gyr; \citealt{leroy:2008}) but in line with the current paradigm that low-mass galaxies are the least efficient at converting their gas into stars \citep[e.g.,][and references therein]{behroozi:2013}. %It is slightly larger than the $\text{SFE}\sim0.001$--0.01 Gyr$^{-1}$ found in earlier chemical evolution studies of the Hercules and Bo\"otes {\sc I} UFDs by \citet{vincenzo:2014}, but this \edit1{\deleted{is}} difference is likely predominantly driven by their assumption of mass-loading factors an order of magnitude smaller than we infer here ($\eta\sim10$ vs.\ 200).

\begin{figure*} 
    \begin{center}
    \includegraphics[width=\textwidth]{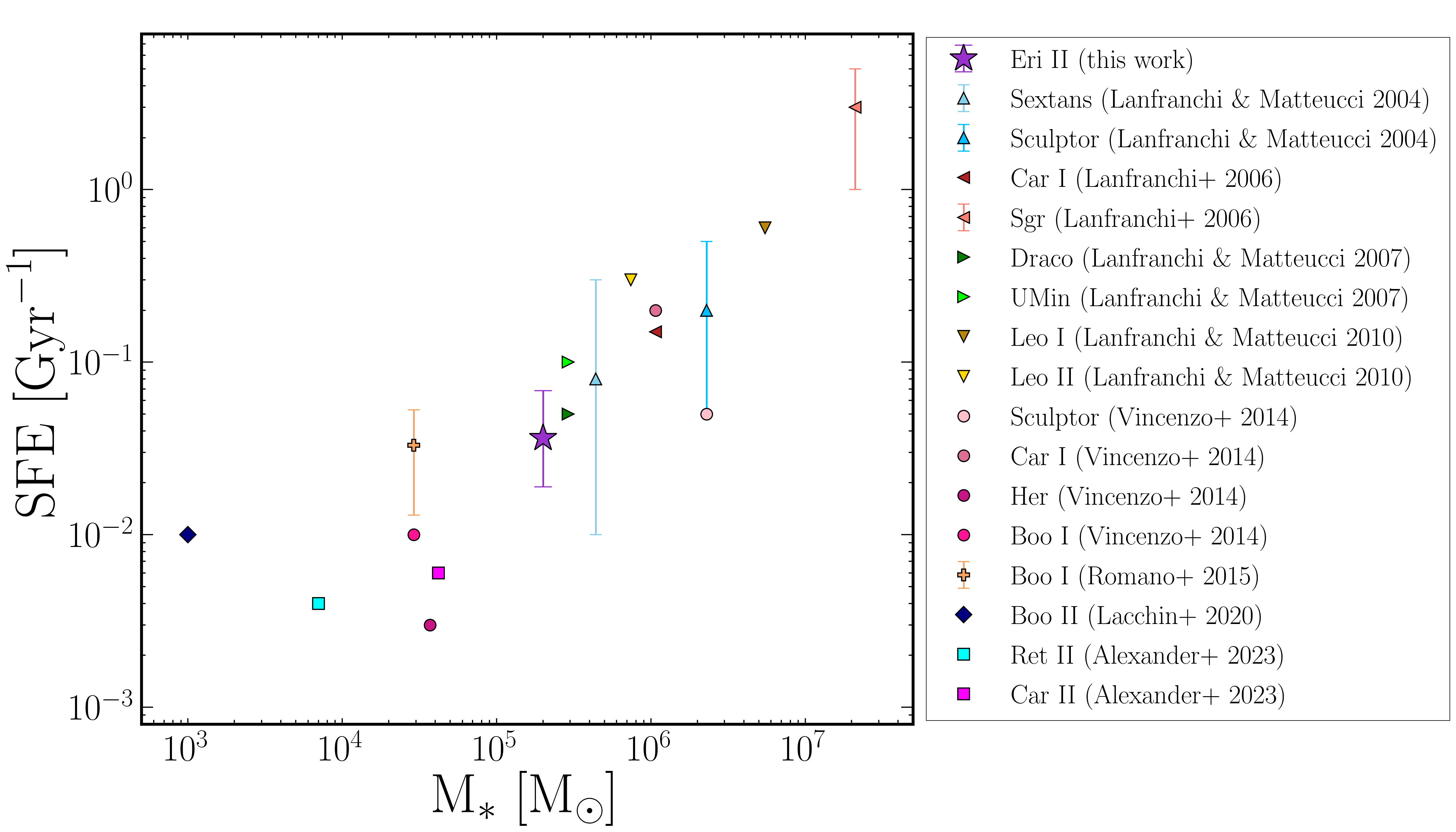}
    \caption{
        The inferred SFE of Eri II (purple star) compared to the SFEs reported by previous chemical evolution studies of LG dwarf galaxies. Though these studies employed a range of chemical evolution models, a clear relationship between galaxy stellar mass and SFE is apparent. The SFE inferred for Eri II is in good agreement with this relationship given its stellar mass.
        \label{fig:SFE_Mstar}
    }
    \end{center}
\end{figure*}

In Figure \ref{fig:SFE_Mstar}, we compare the inferred SFE of \EriII\ with the best-fit SFE reported by previous chemical evolution studies of 13 LG dwarf galaxies spanning a wide range in stellar masses\citep[][]{lanfranchi:2004, lanfranchi:2006, lanfranchi:2007, lanfranchi:2010, vincenzo:2014, romano:2015, lacchin:2020, alexander:2023}. Here, we adopt a stellar mass for \EriII\ of $2\times10^{5}$ M$_\odot$ from \citet{gallart:2021}. Despite the range in chemical evolution models and assumptions adopted in these studies, a clear trend between a galaxy's SFE and its stellar mass is visible. As expected, galaxies more massive than \EriII\ are found to be more efficient at converting gas to stars ($\text{SFE}\sim0.5$--$1.0$ Gyr$^{-1}$), while galaxies less massive than \EriII\ are found to be less efficient ($\text{SFE}\sim0.003$--$0.03$ Gyr$^{-1}$). Given this apparent relationship, the SFE we infer for \EriII\ is in good agreement with expectations given its stellar mass. 

A low SFE like that found for \EriII\ may be indicative that the majority of \EriII's gas is in the atomic phase. Following the reasoning of \citet{johnson:2020} that $\tauSFE=(2~\text{Gyr})(1+M_\text{H{\sc I}}/M_{\text{H}_2})$, we infer that the molecular gas fraction in \EriII\ was only $7.26\pm_{3.46}^{6.41}$\%. This, of course, is only a rough approximation given the assumptions made in our model. It is possible that the molecular gas fraction of \EriII\ changed over its star-forming lifetime, resulting in a time-varying SFE. \citetWAF{}\ find that smooth evolution of \tauSFEtext\ has little impact on chemical evolution tracks if the SFH remains fixed, but we have not yet investigated the low-metallicity regime relevant here. While numerical solutions with time-dependent SFE are feasible, it is not clear what behavior would be appropriate to assume for \EriII, and so we stick with the simplest assumption of constant SFE.

\paragraph{Star Formation History}
We recover the SFH timescale of \EriII\ to be $\tauSFH=0.39\pm_{0.13}^{0.18}$ Gyr. This is slightly shorter than the SFH reported by \citet{gallart:2021}, which informed our choice of prior ($\tau_\text{SFH,prior}=0.7\pm0.3$ Gyr).  The SFH timescale we infer corresponds to a star formation HWHM of $270\pm_{90}^{130}$ Myr -- about half the duration found by \citet{gallart:2021}. 
This supports the hypothesis of \citet{gallart:2021} that the true duration of \EriII's main star formation episode is shorter than they could resolve with their CMD fitting techniques. Assuming star formation commenced immediately, the inferred \tauSFHtext\ implies that \EriII\ had formed $\sim$65\% of its stellar mass by $z\sim11.5$ and $\sim$95\% of its stellar mass by $z\sim5.7$, which would independently confirm that \EriII\ is a relic of pre-reionization era galaxy formation.

While we can place tight constraints on the SFH timescale, the inference of \ttrunctext\ ($t_\text{trunc}=1.37\pm_{0.37}^{0.37}$ Gyr; $z_\text{trunc}=4.40\pm_{0.81}^{1.25}$) remains dominated by the imposed prior. Tests allowing \ttrunctext\ to be unconstrained find no evidence in \EriII's MDF that the SFR truncated abruptly within the first 5 Gyr. That said, we know from \EriII's CMD that there has been effectively no star formation for the last $\sim$13 Gyr. Our inability to provide independent constraints on \ttrunctext\ is not indicative of tension between the MDF and CMD but rather a result of how subtle the impact of truncation is given \EriII's short star formation timescale. In our best-fit model, truncation occurs after $\sim$3.5\tauSFHtext, when the SFR is already quite low -- \EriII's stellar mass would only increase by $\sim$3\% in the absence of truncation. In other words, the inferred exponential suppression of \EriII's SFH is already strong enough that a final, super-exponential truncation is difficult to detect. Constraining a sharp truncation in \EriII's SFH from its MDF would require a larger sample of stars with abundance measurements precise enough to map the high-metallicity tail of the MDF.

\paragraph{Mass-Loading Factor}
We recover a broad but clearly peaked posterior for the galaxy's mass-loading factor, $\eta=195\pm_{43}^{33}$. This means that for every 1 M$_\odot$ of star formation, nearly 200 M$_\odot$ of ISM gas is ejected from the galaxy by SNe feedback. While extreme in comparison to the mass-loading factors of MW-like galaxies ($\eta\sim1$), mass-loading factors of this magnitude are frequently invoked for low-mass galaxies in order to match simulations to empirical scaling relations \citep[e.g.,][]{benson:2003, somerville:2015, mitchell:2020}. State-of-the-art hydrodynamic simulations have also found $\eta\sim100$ for the lowest-mass dwarf galaxies \citep[e.g.,][]{muratov:2015, emerick:2019, pandya:2021}.

\begin{figure} 
    \begin{center}
    \includegraphics[width=\columnwidth]{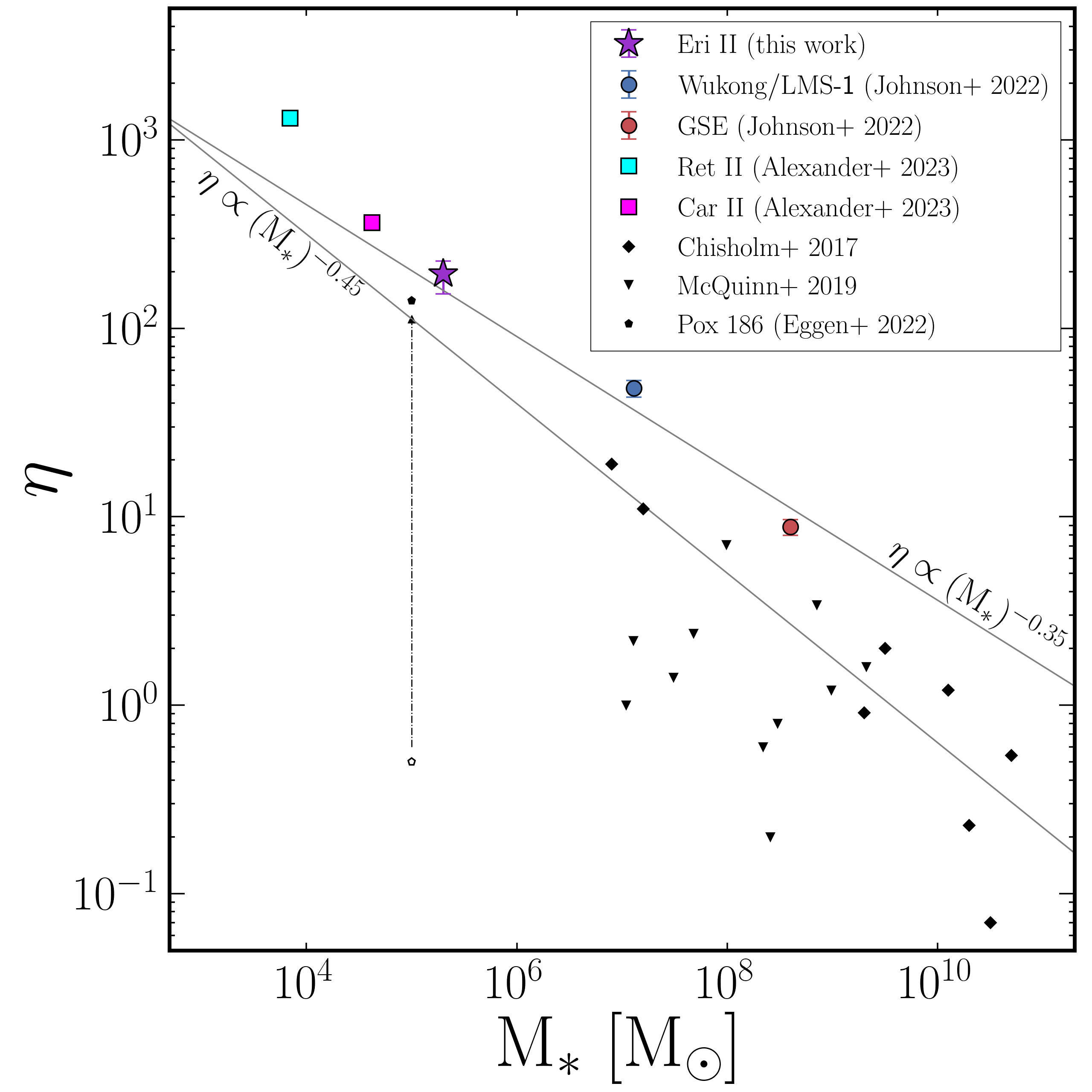}
    \caption{
        The inferred Mass-loading factor of Eri II (purple star) compared to the mass-loading factors inferred by the chemical evolution studies of \citet{johnson:2022b} for Wukong/LMS-1 and GSE (blue and red circles respectively) and \citet{alexander:2023} for Car II and Ret II (magenta and cyan squares respectively) as a function of stellar mass. Mass-loading factors for galaxies observed by \citet{chisholm:2017} and \citet{mcquinn:2019} are included as black diamonds and triangles respectively. The current observed mass-loading factor of Pox 186 and its previous estimated mass-loading factor from \citet{eggen:2022} are represented by the open and filled black pentagons respectively. The scaling found by \citet{muratov:2015} in FIRE-1 simulations indicative of momentum-driven winds ($\eta\propto M_*^{-0.35}$) and the scaling found by \citet{pandya:2021} in FIRE-2 simulations indicative of energy-driven winds ($\eta\propto M_*^{-0.45}$) are included for reference as solid black lines.
        \label{fig:Eta_Mstar}
    }
    \end{center}
\end{figure}

In Figure \ref{fig:Eta_Mstar}, we compare our inferred mass-loading factor for \EriII\ (purple star) to the mass-loading factors inferred from chemical evolution studies of the disrupted dwarf galaxies \textit{Gaia}-Sausage Enceladus (GSE) and Wukong/LMS-1 by \citet{johnson:2022b} using the \texttt{VICE} one-zone chemical evolution model (red and blue circles respectively) and of the UFDs Carina II and Reticulum II by \citet{alexander:2023} using the \texttt{i-getool} inhomogenous chemical evolution model (magenta and cyan squares respectively). In addition, we include measurements of the mass-loading factor of galaxies from \citet{chisholm:2017} and \citet{mcquinn:2019} (black diamonds and triangles respectively). Lastly, we include the dwarf starburst galaxy Pox 186, which was observed by \citet{eggen:2022} to currently have a suppressed mass-loading factor (open black pentagon) due to the efficient removal of gas by earlier SN-driven outflows. By estimating the total amount of gas lost to outflows, \citet{eggen:2022} concluded that the mass-loading factor of Pox 186 was substantially larger during its previous outflow episode (filled black pentagon).

Direct comparison between mass-loading factors measured through direct observational indicators \citep[e.g.,][]{chisholm:2017, mcquinn:2019} and chemical evolution models \citep[][and this work]{johnson:2022b, alexander:2023} is challenging for a number of reasons. For one, the manner in which outflows are parametrized in models frequently do not map directly to the observable quantities being measured. Additionally, the strength of outflows experienced presently at $z=0$ by a galaxy of a given mass may not be representative of the outflows experienced at high redshift by galaxies of a similar mass. Nevertheless, our result for \EriII\ is in good qualitative agreement with the observed trend that less massive galaxies have stronger outflows.

Two scaling relationships have historically been invoked for relating the mass-loading factor of a galaxy to its stellar velocity dispersion, $\sigma$ (as a proxy for its mass). In the physical scenario of momentum-driven winds governed by radiation pressure, the mass-loading factor scales as $\eta\propto V_{c} ^{-1}$ \citep{murray:2005}. This scaling has been argued for by \citet{finlator:2008} and \citet{peeples:2011} based on the observed mass-metallicity relationship. Based on results from the FIRE-1 cosmological zoom-in simulation \citep{hopkins:2014}, \citet{muratov:2015} found a scaling relationship between the mass-loading factor and galactic stellar mass of $\eta=3.6(M_*/10^{10}~\text{M}_\odot)^{-0.35}$, which is in good agreement with the expectations given a momentum-driven wind scaling. Alternatively, in the physical scenario of energy-driven winds from SNe, the mass-loading factor scales as $\eta\propto V_{c} ^{-2}$ \citep{chevalier:1985}. This scaling has been argued to be more important in low-mass galaxies with $\sigma<75$ km s$^{-1}$ \citep[e.g.,][]{murray:2005, murray:2010, hopkins:2012, dave:2013}. Furthermore, \citet{pandya:2021} found a steeper scaling relationship in the updated FIRE-2 simulations \citep{hopkins:2018} of $\eta=0.6(M_*/10^{10}~\text{M}_\odot)^{-0.45}$, which is more characteristic of the energy-driven wind scaling.

For reference, we have included both the momentum-driven scaling ($\eta\propto M_*^{-0.35}$) of \citet{muratov:2015} and the energy-driven scaling ($\eta\propto M_*^{-0.45}$) of \citet{pandya:2021} in Figure \ref{fig:Eta_Mstar}. The normalization of the chemistry-based mass-loading factors, including \EriII, is more in-line with the findings of \citet{muratov:2015}. However, their scaling with stellar mass is marginally closer to that of \citet{pandya:2021}, just offset by a small factor to larger values. A larger sample of galaxies, especially at the lowest masses, is required before these measurements can discriminate between these two physical outflow scenarios.

While smaller mass-loading factors ($\eta\lesssim50$) are not prohibited by the model, they are $2\sigma$ disfavored and would require longer SFE and SFH timescales. Allowing for direct ejection of SN products (e.g., letting $\fret<1$) has the potential to temper large mass-loading factors, but preliminary tests suggest that 1) mass-loading values of $\eta\sim100$ are still preferred and 2) lower mass-loading factors require both low retention fractions ($\fret\sim0.3$) and higher SFEs ($\tauSFE\sim1.0$) -- see the Metal-Loading model in Section \ref{sec:alt_model_fits}. As with the \tauSFEtext, $\eta$ could in principle vary with time as \EriII's dark matter halo grew and its potential well deepened, though large changes in $\eta$ over the duration of \EriII's star-forming lifetime are disfavored by its small \tauSFHtext. We leave investigation of a time-dependent $\eta$ for future study.

\subsection{Alternative Model Fits}
\label{sec:alt_model_fits}
\begin{figure*}
    \begin{center}
    \includegraphics[width=\textwidth]{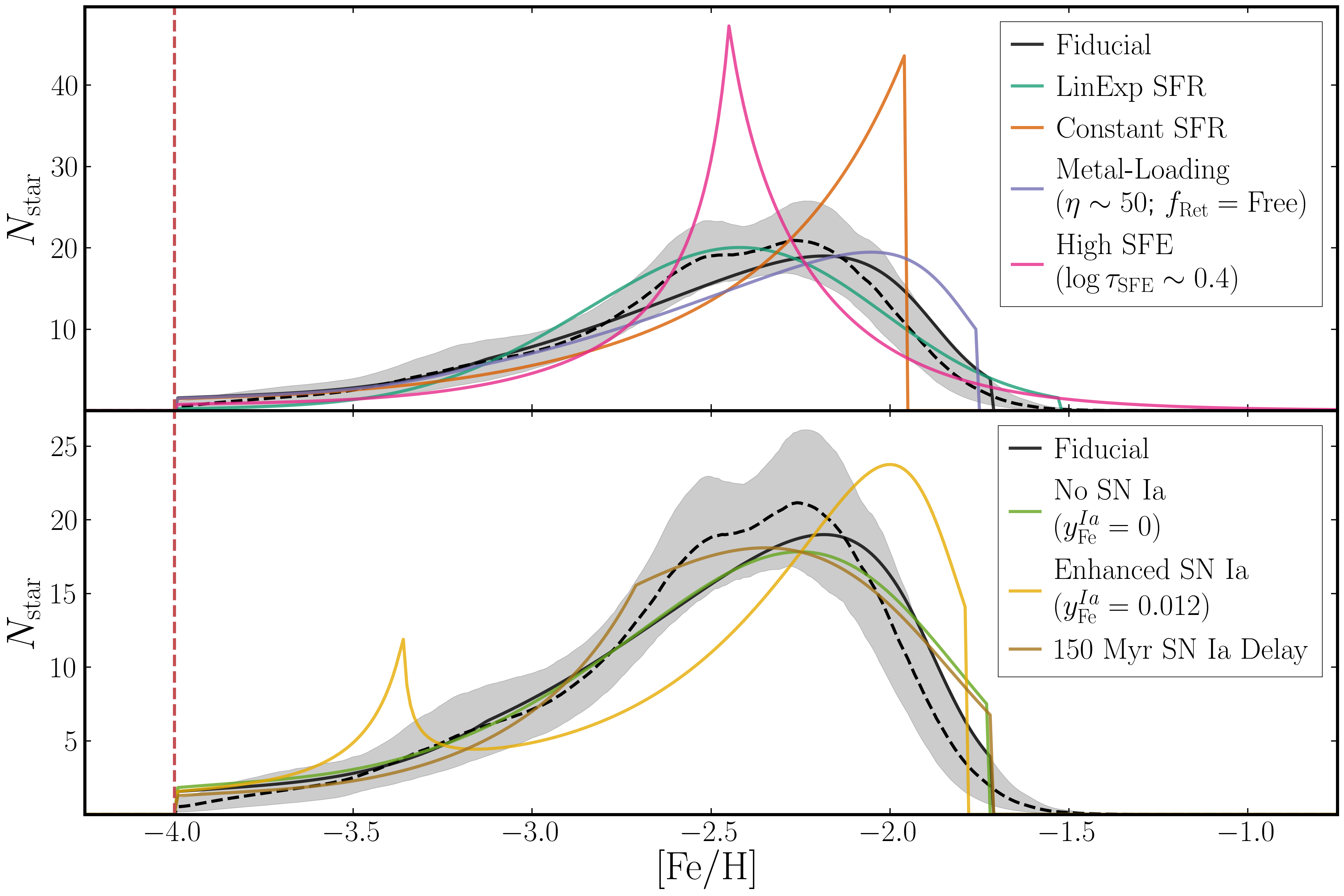}
    \caption{
        Best-Fit MDFs generated with the alternative models described in Section \ref{sec:alt_models} (colored lines) compared to the Fiducial model's best-fit MDF (solid black line) and latent [Fe/H] distribution (dashed black line and grey shaded region). 
        \label{fig:Alternative_Model_Comparison}
    }
    \end{center}
\end{figure*}

Here we present the results of fitting the \EriII\ MDF with the alternative models described in Section \ref{sec:alt_models}. Median and 16th- and 84th-percentiles of the marginalized posteriors are presented alongside the fiducial best-fit values in Table \ref{tab:best_fit_par}. In Figure \ref{fig:Alternative_Model_Comparison}, we compare the MDF predicted for these median posterior values of each alternative model (colored lines) to that of the Fiducial model (black line) and its latent \FeHtext\ distribution (gray dashed line and shaded region). We refer to these predicted MDFs as the ``best-fit" for each model, though strictly speaking the model with the highest posterior probability does not have exactly the median posterior values of each parameter. 

It is not entirely fair to judge the quality of the alternative model fits to the fiducial latent MDF, as each model may predict a different underlying distribution. In practice, however, we find that latent MDF of most models is quite similar to the fiducial case. The two exceptions to this are the latent distributions of the Constant SFR and High SFE models, which are more negatively skewed and centrally peaked respectively. While not statistically prohibited given the wide posteriors of the CaHK measurements, such underlying MDFs would be unusual in comparison to the MDFs observed in other dwarf galaxies using more precise spectroscopic abundances. 

One quantitative metric for judging the goodness-of-fit of each model relative to the Fiducial model is the Bayes factor (BF). The BF is defined as the ratio of the Bayesian evidence of each model, which expresses the posterior probability of one model relative to the Fiducial model under the a priori assumption that both models are equally probable. A $\text{BF}<1$ indicates that the Fiducial model is more probable while a $\text{BF}>1$ indicates the alternative model is more probable. Using the Bayesian evidence estimated by \texttt{pocomc}, we calculate the BF for each model; we present these in the right-most column of Table \ref{tab:best_fit_par}.

In general, we find BFs $<$ 1, indicating that the Fiducial model is preferred. However, we caution that the BF is inherently sensitive to choices in priors, so BFs of $\mathcal{O}(1)$ should not be over-interpreted. Nevertheless, because the BFs of the High SFE and Enhanced SN Ia are $\ll$1, we can be fairly confident that those models are disfavored. With a BF of $\sim$0.2, the Constant SFR model is also disfavored but to a more modest degree. One could argue that many of the alternative models require specific values for model parameters, making them a priori less likely. In this case, the BF we calculate would overestimate the probability of the alternative models relative to the Fiducial model. Definitively ruling out alternative models with $\text{BF}\sim1$ would require additional observational constraints, including a larger sample of stars, more precise \FeHtext\ measurements, and/or measurements of \alphaFetext\ (see Section \ref{sec:additional_model_predictions}). 

We discuss each alternative model and their MDF predictions below. 

\paragraph{Linear-Exponential SFR Model} 
    Despite the distinctly different functional form of the Linear-Exponential SFR model, we find that it provide a fit to the data which is just as good (if not slightly better) than that of the Fiducial model, while inferring very similar values for \tauSFEtext, \ttrunctext, and $\eta$. Though the MDF predicted by this model is more symmetric, peaking at a slightly lower metallicity ($\FeH \sim -2.5$) and extending to a higher metallicity ($\FeH \sim -1.5$), the dispersion and mean metallicity of the MDF are in good agreement with the Fiducial model's predictions. The Bayes factor slightly favors this fit over the Fiducial, but only marginally so. Discriminating between these two models would require a larger sample of high quality stellar metallicities than is currently available. 

Our primary conclusion from this model is that our constraints on \tauSFEtext\ and $\eta$ from the fiducial model are \textit{not} sensitive to the assumed SFH at early times. In particular, our findings are insensitive to the Fiducial model's assumption that \EriII's gas reservoir was already in place at the onset of star formation, instead of growing rapidly through gas accretion as it does in the Linear-Exponential SFR model (see Section \ref{sec:initial_final_conditions}).

\paragraph{Constant SFR Model} 
The Constant SFR model predicts an MDF that is rapidly increasing until it sharply truncates at the high-metallicity end. Such an MDF is atypical for dwarf galaxies in the Local Group, which universally exhibit MDFs that turn over at the high-metallicity end \citep[e.g.,][]{kirby:2011a}. The outflow mass-loading required to achieve the observed metallicity range is an order-of-magnitude below that of the Fiducial model because a constant SFR requires rapid continuing gas accretion that dilutes the metal production from stars.  Relaxing the prior on \ttrunctext\ results in an earlier truncation ($\sim$0.6 Gyr) and a slightly smaller \logtauSFEtext\ ($\sim$1.7), but the shape of the predicted MDF and the inferred mass-loading factor remain largely the same. Although the predicted MDF shape is radically different from the \citetFu{} histogram and from the latent MDF of the Fiducial model, the Bayes Factor only mildly disfavors the Constant SFR mode. The uncertainties in the CaHK \FeHtext\ values are large enough that a sharply truncated MDF is not in large tension with the measurements. More precise \FeHtext\ values for the highest metallicity stars in \EriII\ would be needed to decisively distinguish the Fiducial and Constant SFR models.

\paragraph{Metal-Loading Model} 
We find that forcing a lower mass-loading factor ($\eta\sim50$) is able to produce realistic fits to the \EriII\ MDF if the retention fraction of SN products is low ($\fret\sim0.3$) -- that is, 70\% of the metals produced by SNe are directly ejected from the galaxy. The covariance we find between $\eta$ and \frettext\ in models where both are free implies that $(\fret,~\eta) = (0.3,~50)$ is roughly consistent with the mass-loading factor of $\eta\sim200$ inferred in the Fiducial model when $\fret=1$. Conversely, this covariance implies that to produce a similar MDF with an even smaller mass-loading factor of $\eta\sim10$ would require a retention factor of only 10\% (and also a substantially lower SFE timescale of $\logtauSFE\sim0.5$). 
While the best-fit MDFs of the Fiducial and Metal-Loading models are both generally consistent with the data (the former more-so than the latter), they do produce qualitatively different MDFs. The Metal-Loading model predicts a slightly more skewed MDF with a higher metallicity peak and a truncated metal-rich tail. These differences result from the fact that the Metal-Loading model is always losing a significant fraction of the metals produced, while the Fiducial model only experiences significant metal losses once the ISM metallicity approaches its final value. Discriminating between these two models (i.e., breaking the degeneracy between $\eta$ and \frettext) can therefore be accomplished by acquiring more precise \FeHtext\ measurements of stars in the high-metallicity end of the MDF.

\paragraph{High SFE Model}
When we force a short SFE timescale ($\logtauSFE\sim0.4$), we find that the best-fit MDF is sharply peaked at $\FeH\sim-2.4$ with roughly exponential tails on either side. The peak of this distribution corresponds to the CC equilibrium Fe-abundance that the model would evolve to in the absence of SN Ia (see \citetWAF{} for derivation): 
\begin{equation}
    \label{eq:FeHCCEq}
    \FeH_\text{eq}^\text{cc} = \log_{10}\left(\frac{\yFeCC}{Z_{\text{Fe},\odot}(1 + \eta - r - \tauSFE/\tauSFH)}\right).
\end{equation}
For a galaxy with an exponentially declining SFH, this equilibrium value is approached on the ``harmonic difference timescale" set by \tauSFHtext\ and the gas depletion timescale $\tau_\text{dep}$,
\begin{equation}
    \label{eq:tauCCEq}
    \tau_\text{Fe,eq}^\text{cc} \equiv \left(\tau_\text{dep}^{-1} - \tauSFH^{-1}\right)^{-1},
\end{equation}
where the depletion timescale is defined to be 
\begin{equation}
    \label{eq:tauDep}
    \tau_\text{dep} = \frac{\tauSFE}{1 +\eta - r}.
\end{equation}
Equation \ref{eq:tauCCEq} can be equivalently expressed as
\begin{equation}
    \label{eq:tauCCEq2}
    \tau_\text{Fe,eq}^\text{cc} = \frac{\tauSFE}{1 + \eta - r - \tauSFE/\tauSFH}.
\end{equation}

While both this model and the Fiducial model have $\FeH_\text{eq}^\text{cc}\sim-2.4$, the short \tauSFEtext\ imposed here results in a substantially smaller $\tau_\text{Fe,eq}^\text{cc}$ ($\sim$30 Myr vs. $\sim$200 Myr), which is shorter than the 50 Myr minimum time delay for SN Ia. As a result, the model quickly evolves to the equilibrium metallicity where it forms stars until the onset of SN Ia at which point the model evolves to higher metallicity. The sharp decline in the MDF above $\FeH_\text{eq}^\text{cc}$ is due to the short SFH timescale (0.08 Gyr) inferred for this model -- by the time SN Ia begin increasing the metallicity, the rate of star formation is rapidly declining. In contrast, the Fiducial model's longer $\tau_\text{Fe,eq}^\text{cc}$ is sufficiently long that SNe Ia begin contributing to Fe production before $\FeH_\text{eq}^\text{cc}$ is reached, resulting in a smoother MDF with no sharp peaks. The High SFE model is strongly disfavored by the Bayes factor ($\sim$0.002), and its extremely short \tauSFHtext\ appears physically implausible.

\paragraph{Longer SN Ia Delay} 
Within the uncertainties of the latent MDF, we find that a model assuming a minimum SN Ia time delay of 0.15 Gyr provides a fit that is roughly as good as that of the Fiducial model. With this longer time delay, we infer a SFH timescale that is slightly larger and a mass-loading factor that is slightly smaller than the Fiducial model. The kink in the evolutionary track at $\FeH\sim-2.75$ corresponds to the onset of SN Ia enrichment (a milder version of the sharp transition found in the High SFE model). In principle one could distinguish the $t_D=0.05$ and 0.15 Gyr scenarios from the different shapes of the predicted MDFs, but this would require a larger sample of high quality stellar metallicities than is currently available.

\paragraph{No SN Ia} 
We find that a model with no SN Ia enrichment can reproduced \EriII's MDF reasonably well with only slight differences from the MDF of the Fiducial model. Like the Metal-Loading model, this model is most distinguishable in the high-metallicity tail. While this model infers values for the SFE and SFH timescales and \ttrunctext\ that are consistent with the Fiducial model, it requires a mass-loading factor that is $\sim$2.5 times smaller because the total Fe yield is lower. Unlike the other best-fit models, this model has a SFH timescale that is \textit{shorter} than the depletion timescale $\tau_\text{dep}$ (Equation \ref{eq:tauDep}), 0.45 Gyr vs.\ 0.52 Gyr, which is necessary in the absence of SN Ia enrichment to avoid forming an MDF peaked at $\FeH_\text{eq}^\text{cc}$. Physically achieving $\tauSFH<\tau_\text{dep}$ would require the removal of gas from the galaxy by a process not associated with star formation. If this is indeed the case for \EriII\, then reionization-driven photo-evaporation might be responsible for the removal of gas, though this would require additional investigation. 

While the Bayes factor of this model is marginally larger than the Fiducial model, we caution that this alone does not indicate that the No SN Ia model is better. The BF is only informative insofar as the two models are equally likely a priori. The scenario considered here, in which no SN Ia contributed in any part to the Fe enrichment of the stars in our sample, is highly improbable given the expectation that $\sim$200 SN Ia should have occurred in a galaxy of \EriII’s mass. The No SN Ia model could be easily distinguished from the Fiducial model with measurements of \alphaFetext\ ratios, which should remain elevated in the absence of SN Ia enrichment (see Section \ref{sec:MgFe_predictions}).

\paragraph{Enhanced SN Ia} 
Increasing the SN Ia yield by a factor of 10, as a SN Ia rate $\propto Z^{-0.5}$ would imply, results in an MDF with  a higher and narrower peak close to the eventual sharp truncation, as well as a secondary low-metallicity peak. The high \yFeIatext\ forces a  high $\eta$, which in turn leads to a short depletion time $\tau_\text{dep}\sim30$ Myr. As with the High SFE model, this low-metallicity peak is the result of the model evolving to its equilibrium CC Fe abundance (in this case $\FeH_\text{eq}^\text{CC}\sim-3.25$) before the commencement of SN Ia at $t_D=50$ Myr. The Enhanced SN Ia model is disfavored by the Bayes factor ($\sim$0.02). 

The discrepancy on the high-metallicity end of the MDF is alleviated if we allow for $\fret<1$, which compensates by decreasing the effective SN yield. Indeed, if we set $\fret=0.1$ for SN Ia but $\fret=1$ for CCSN, then this model is equivalent to the Fiducial model, with direct SN Ia metal loss exactly cancelling the higher yield. However, there is no obvious reason to have a high retention fraction for CCSN but a low retention fraction for SN Ia. We do find that adopting a single $\fret=0.1$ produces an MDF in better agreement with the Fiducial model. While the best-fit mass-loading factor in this case is similar to that of the Fiducial model ($\eta\sim200$), the inferred SFE timescale is substantially lower ($\logtauSFE\sim0.7$). If we allow \frettext\ to be free, the model prefers larger retention fractions and mass-loading factors: $\fret\sim0.6$ and $\eta\sim700$ (though these parameters, as always, are very degenerate). In all of these permutations, the peak around $\FeH_\text{eq}^\text{CC}$ remains.

\section{Discussion}
\label{sec:discussion}

\subsection{Physical Interpretation of the Model}
\label{sec:discussion_par}
Our Fiducial model achieves a good match to the observed MDF with physically plausible values of its four evolutionary parameters, $\tauSFE=27.5$ Gyr, $\tauSFH=0.39$ Gyr, $\ttrunc=1.37$ Gyr, and $\eta=194$. As previously discussed, a low SFE (large $\tauSFE$) is characteristic of low-$M_*$ dwarfs, and the high $\eta$ value is consistent with scaling relationships from numerical and analytic models extrapolated to the low mass of \EriII. The $e$-folding timescale for star formation is consistent with direct estimates of the SFH \citep{gallart:2021}, but the value of $\ttrunc$ is not independently well constrained by the MDF data.

\begin{figure} 
    \begin{center}
    \includegraphics[width=\columnwidth]{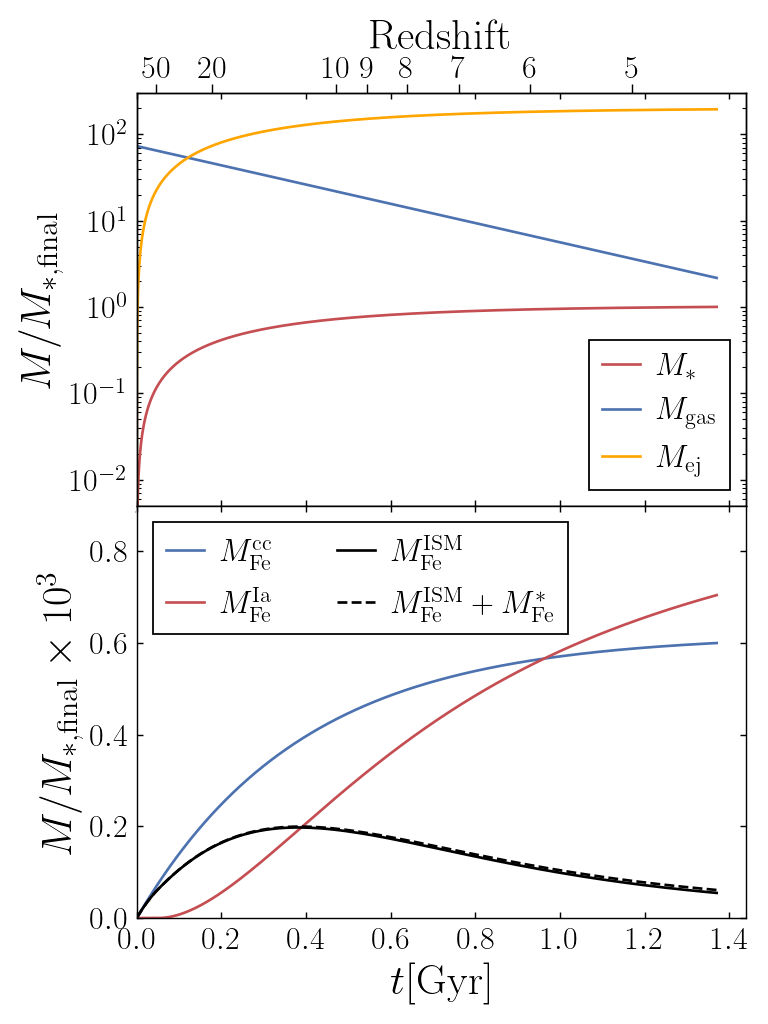}
    \caption{
        Top: The stellar mass (red), gas mass (blue), and cumulative ejected gas mass (orange) of the best-fit model as a function of time. The star-to-gas ratio is very low at early times but evolves close to unity by the time star formation ceases. At the end of the simulation, the model has lost roughly $100\times$ its stellar mass in gas outflows.
        Bottom: The mass of Fe in the ISM (solid black) and the cumulative mass of Fe produced by CCSNe and SN Ia (blue and red respectively) for the best-fit model as a function of time. At early times, CCSNe dominate the production of Fe, while at late times the total contribution of CCSNe and SN Ia are roughly equal. While the amount of Fe in the ISM is slowly decreasing for $t>0.3$ Gyr, the rapidly decreasing gas supply (top panel) results in a monotonically increasing gas-phase metallicity. The mass of Fe in stars is negligible as can be seen from the dashed black line, which shows the combined mass of Fe in both the ISM and long-lived stars.
        \label{fig:Fiducial_Physical_Properties}
    }
    \end{center}
\end{figure}

Figure \ref{fig:Fiducial_Physical_Properties} elucidates the evolution of the best-fit Fiducial model. Because of the short $\tauSFH$, the model has already formed $\sim$65\% of its stellar mass by $t=0.4$ Gyr ($z=11.3$). The gas mass, $M_{g} = \tauSFE\dot{M}_*$, follows the same exponential decline as the SFH, given the assumption of a constant $\tauSFE$. Through most of the model's evolution the stellar mass fraction $M_*/(M_*+M_g)$ is $\ll$1, though by the end it has risen to 0.5. However, because the value of $\eta$ is so high, the mass of gas ejected from the galaxy exceeds the mass remaining in the ISM at all times $t \gtrsim 0.1$ Gyr. The model has vigorous ongoing gas accretion that fuels continuing star formation despite the strong outflow, with an infall rate $\dot{M}_\text{inf} \approx (\eta-\tauSFE/\tauSFH)\dot{M}_* \approx 120\dot{M}_*$ (see Equation \ref{eq:mdotinfall}). An exponential SFH requires a non-zero gas mass at $t=0$. In the Fiducial model, the mass of gas accreted exceeds this initial mass for $t \gtrsim 0.35$ Gyr. 

The bottom panel of Figure \ref{fig:Fiducial_Physical_Properties} tracks the Fe mass budget. In the Fiducial model, the total Fe produced by CCSN and SN Ia is nearly equal over the life of the galaxy. However, CCSN enrichment dominates the early evolution, and by late times the enrichment rate from SN Ia greatly exceeds that from CCSN, as one can see by comparing the slopes of the blue and red curves. These conclusions rely on our adopted values of \yFeCCtext\ and \yFeIatext; as shown in Figure \ref{fig:Alternative_Model_Comparison}, the MDF can be reasonably well reproduced even in a model with no SN Ia enrichment. For $t > 0.4$ Gyr, the mass of Fe ejected from the galaxy exceeds the mass remaining in the ISM by a substantial factor. The Fe mass in stars is small compared to that in the ISM because the star-to-gas mass ratio is low and because the mean metallicity of stars is always lower than the ISM metallicity.

Although the SFE timescale is long, the gas depletion time (Equation \ref{eq:tauDep}) is short because of the high $\eta$, $\tau_\text{dep}\sim0.14$ Gyr. As discussed by \citetWAF{}, obtaining an MDF that turns over rather than peaking sharply requires a rapidly declining SFH so that newly produced metals are deposited in a dwindling gas supply, resulting in high ISM metallicity at late times when only a small number of stars are produced.  In the case of the Fiducial model, $\tauSFE=0.39$ Gyr is still significantly longer than $\tau_\text{dep}$, but it is shorter than the characteristic enrichment time for SN Ia (roughly 1.5 Gyr), so the SN Ia enrichment drives the MDF turnover.
In the $\yFeIa=0$ model, by contrast, the MDF turnover arises because the depletion time is longer, and with $\tauSFH\sim\tau_\text{dep}$ the model approaches the leaky box scenario, which was shown to provide a reasonable fit to the \EriII\ MDF by \citetFu{}. Our Fiducial model is still rather far from this limit.

We can summarize the physical properties of the Fiducial model as follows. It begins with an initial gas mass $M_g \sim 1.6\times 10^7$ M$_\odot$
and accretes gas vigorously but at an exponentially declining rate. Only a small fraction of the accreted gas forms into stars because the SFE is low and because feedback from star formation drives ISM gas out of the galaxy's shallow potential well with a high mass-loading factor $\eta \sim 200$.  Fe enrichment is dominated by CCSN at early times and by SN Ia at later times, with the two channels producing similar total amounts of Fe over the life of the galaxy. However, more than 90\% of the Fe produced by the stars is ejected from the galaxy. This low metal retention is the main reason for the galaxy's low final metallicity, not the truncation of star formation. The turnover in the MDF arises because Fe from SN Ia is deposited in a dwindling gas supply, enabling a small fraction of stars to form at relatively high [Fe/H]. The exponentially declining SFH arises
because gas accretion does not keep up with gas losses from the feedback-induced galactic wind. Star formation ceases abruptly at $t=\ttrunc\sim1.4$ Gyr ($z\sim4.3$), presumably because reionization evaporates the galaxy's remaining gas supply.

\subsection{Additional Model Predictions}
\label{sec:additional_model_predictions}

\subsubsection{Predictions of \MgFetext}
\label{sec:MgFe_predictions}
While we do not consider \MgFetext\ in our fit (there being presently no stars in \EriII\ with \alphaFetext\ measurements of any kind), we can use our model to make predictions of the \MgFetext\ evolution in \EriII\ that next-generation spectroscopic facilities will soon be able to test \citep[e.g.,][]{sandford:2020b}. In principle, measurements of \MgFetext\ for even a few stars in our sample should provide tighter constraints on our posteriors. 

\begin{figure*} 
    \begin{center}
    \includegraphics[width=\textwidth]{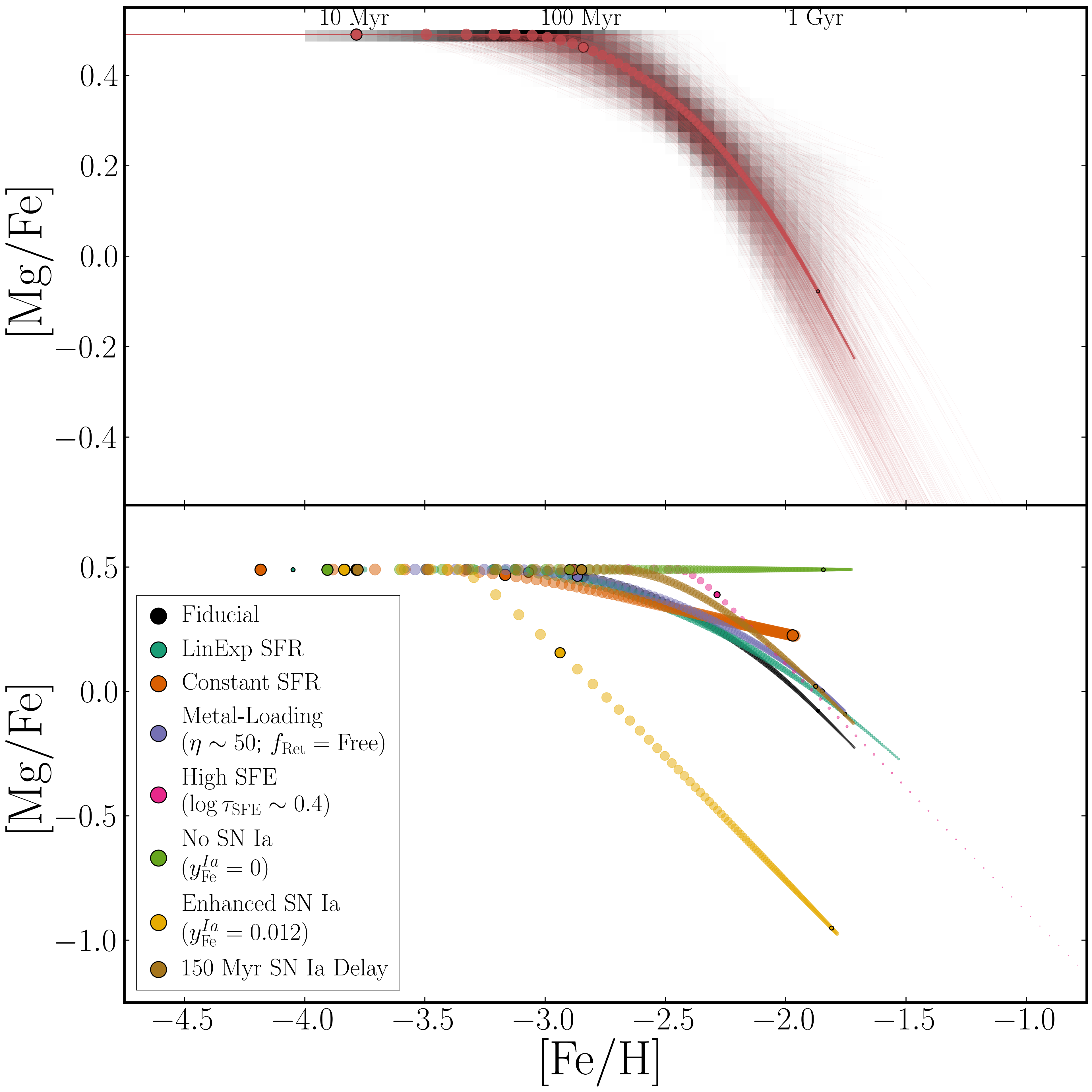}
    \caption{
         Top: [Mg/Fe]-[Fe/H] distribution of stars (gray-scale histogram) predicted by models sampled from our posterior distribution (thin red lines). The evolution of the best-fit Fiducial model in steps of 10 Myr is depicted in red circles; the size of the marker is proportional to the SFR at each step.
         Bottom: The evolution of the best-fit alternative models (colored circles) compared to the Fiducial model's evolution (black circles) following the same plotting convention as the top panel. While many of the alternative models produce MDFs similar to the Fiducial model, they predict quite distinct [Mg/Fe]-[Fe/H] distributions.
        \label{fig:AlphaFe-Fe_PPC}
    }
    \end{center}
\end{figure*}

In Figure \ref{fig:AlphaFe-Fe_PPC} (top), we display the distribution of stars in \alphaFetext-\FeHtext\ space that is compatible with our fiducial \EriII\ model (gray-scale histogram). This distribution is generated by sampling 60 stars from each of the 1000 randomly drawn posterior predictions included in Figure \ref{fig:Fiducial_Model_Fit}. Evolutionary tracks of these models are also included here as thin red lines. Red circles depict 10 Myr snapshots of the best-fit model, where the size of the marker is proportional to the relative SFR at that time; time-steps corresponding to 10 Myr, 100 Myr, and 1 Gyr are outlined in black.

As described in the Section \ref{sec:model}, the low-metallicity plateau of $\MgFe\sim0.5$ is produced by design given our adopted CC Mg and Fe yields. The turn-over or ``knee" in the \MgFetext-\FeHtext\ distribution occurs at $\FeH\sim-3.0$, which is the metallicity of the model at $t=50$ Myr when SN Ia begin to contribute to Fe production. The SN Ia Fe yield was set to $\yFeIa=0.0012$ such that a MW Disk-like model evolves to $\FeH\approx\MgFe\approx0$ at late times. Unsurprisingly, the evolution of a UFD-like model presented here only evolves to $\FeH\sim-1.75$, but it reaches sub-solar Mg abundances of $\MgFe\sim-0.2$ because of the short \tauSFHtext. \MgFetext\ measurements in UFDs are sparse and uncertain, especially for $\FeH\lesssim-3$, but our predictions are generally consistent with observations \citep[e.g.,][and references therein]{simon:2019}.

Though the locus of possible \MgFetext-\FeHtext\ falls relatively tightly around the best-fit Fiducial model, the spread in the posterior predictions is substantial both in the location of the knee and the final \alphaFetext. This suggests, similar to the comparison of alternative models in Figure \ref{fig:Alternative_Model_Comparison}, that the more metal-rich stars in \EriII\ hold increased constraining power. Fortunately, these are also the stars for which spectroscopic measurements should be (comparatively) easier. That being said, measuring \MgFetext\ in stars with $\FeH<-3.0$ will provide valuable constraints on the CC SN yields that determine the high-\MgFetext\ plateau.

The constraining power of \MgFetext\ measurements is further exemplified in the bottom panel of Figure \ref{fig:AlphaFe-Fe_PPC}, where the \MgFetext-\FeHtext\ evolution of the alternative models are compared to that of the Fiducial model. The No SN Ia model (green) is easily distinguishable from the Fiducial model (black) because without SN Ia enrichment \MgFetext\ remains elevated. Meanwhile the Enhanced SN Ia model (yellow) is distinguishable for the opposite reason because the extra SN Ia enrichment drives \MgFetext\ lower faster. The low SFE of the Constant SFR model (purple) leads to SN Ia decreasing \MgFetext\ at lower \MgFetext\ than other models, while its low mass-loading factor means more of the CCSN products produced early are retained, keeping \MgFetext\ from decreasing as steeply. In the High SFE model (pink), \FeHtext\ evolves much more rapidly so the knee occurs at higher metallicity, but given the short \tauSFHtext\ inferred for this model, few stars are formed at lower \MgFetext. The differences predicted by the Longer SN Ia Delay model (brown) and the Metal-Loading model (orange) compared to the Fiducial model are smaller. The longer time delay before SN Ia start shifts the \MgFetext-\FeHtext\ track to higher metallicities, while the direct loss of metals makes SN Ia slightly less effective at decreasing \MgFetext\ at late times. Once again, precise abundance measurements of stars at the high-metallicity end will provide the best opportunity to discriminate between these models.

\subsubsection{Ultra Metal-Poor Stars}
\label{sec:UMP}
Our hierarchical Bayesian framework enables us to recover the posterior distribution of the latent \FeHtext\ for each star in our sample. We caution, however, that these inferred values are influenced by the model MDF and represent the ``true" \FeHtext\ of each star only insofar as the model represents the true MDF of \EriII. Further, because a truncation of the MDF and \FeHtext\ priors below $\FeH < -4$ is imposed by the limitations of the stellar grid used in the CaHK measurements, we cannot recover the metallicity of a star to be more metal-poor than $\FeH < -4$ even if such a star was in our sample. That being said, our framework does allow us to compare the relative probability that each star in our sample was drawn from the un-truncated MDF and the redistributed metal-poor tail of the MDF (see Figure \ref{fig:Example_Model_MDF}). By doing this for the entire posterior sample, we can infer the probability that each star is truly an ultra-metal poor star with $\FeH<-4$.

\begin{figure*} 
    \begin{center}
    \includegraphics[width=\textwidth]{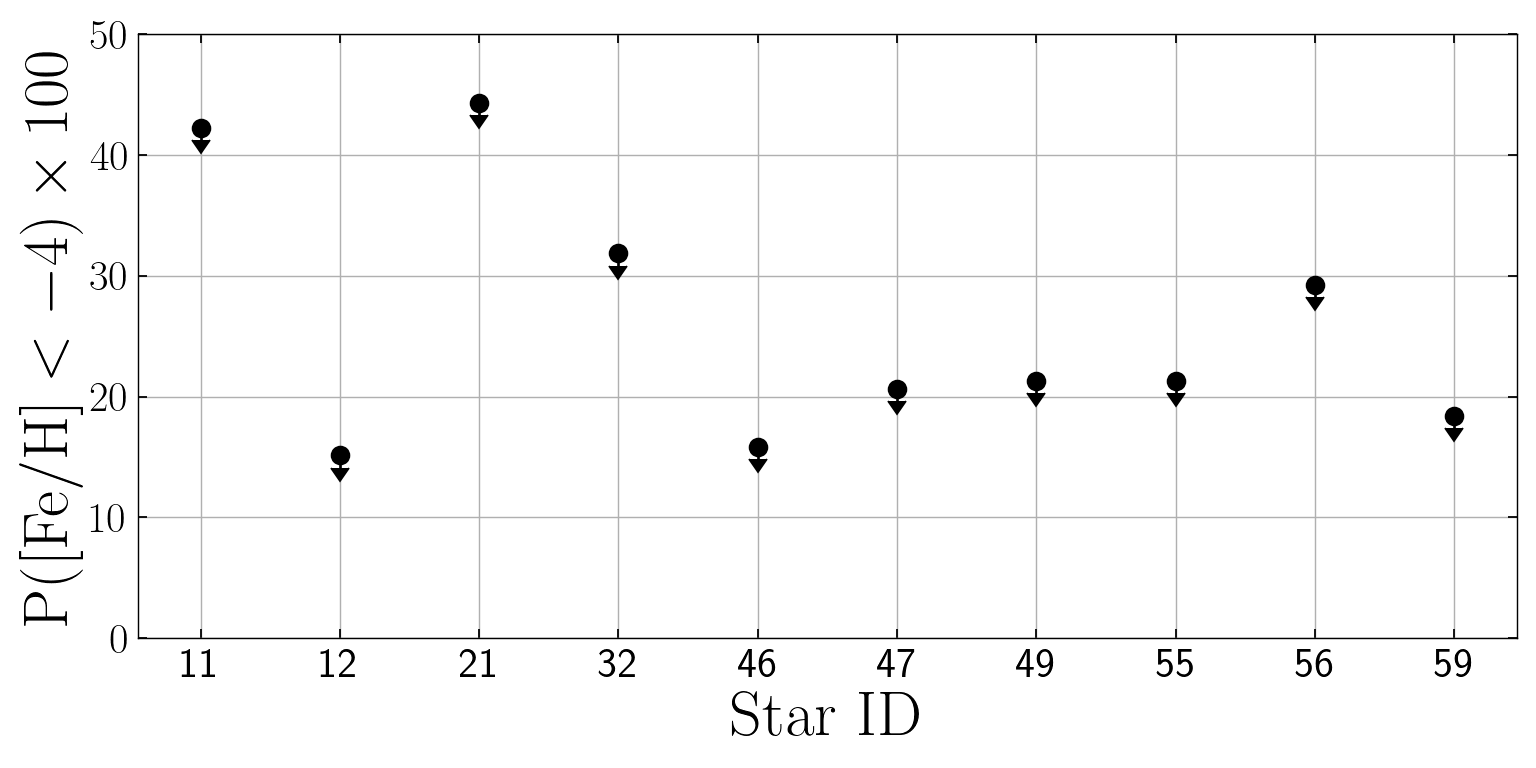}
    \caption{
        95\% upper limits on the inferred probability of metallicities below $\text{[Fe/H]}<-4$ for the highest probability stars in our sample. The remaining 50 stars have probabilities $<$15\%. Stars 11 and 21 were previously identified as extremely metal-poor ($\text{[Fe/H]}<-3$) candidates by F22.
        \label{fig:UMP_Probability}
    }
    \end{center}
\end{figure*}

Figure \ref{fig:UMP_Probability} shows the probability of being an ultra-metal poor star for the 10 highest probability stars. We find that the posterior distribution on $P(\FeH < -4)$ is consistent with zero for every star, strongly disfavoring the presence of any UMP star in our sample. This result suggests that no pre-enrichment of \EriII's gas supply or metallicity floor is necessary to explain the dearth of UMP stars, though a larger sample of stars is necessary to conclusively rule out these scenarios. We place 95\% upper limits on the ultra-metal poor probability of each star and find 10 stars with upper limits greater than 15\%. Stars 11 and 21 have upper limits greater than 40\% and were previously identified as extremely metal-poor ($\FeH<-3$) candidates by \citetFu{}. Searches for UMP stars in \EriII\ should prioritize this subset of our sample for spectroscopic follow-up in order to confirm their true metallicity.

\subsection{Potential Limitations of a One-Zone Model}
\label{sec:limitations}
The analytic solutions employed in our analysis require several idealizations, such as constant values of \tauSFEtext\ and $\eta$. The most important idealization is the one-zone framework itself, i.e., the assumption that the star-forming ISM can be treated as a single, fully-mixed gas reservoir with abundances that evolve in time but do not vary with position. Our key finding is that the Fiducial model reproduces the observed MDF of \EriII\ with parameter values that appear plausible on empirical and theoretical grounds. More complicated models for the chemical evolution of \EriII\ are certainly possible, but they are not required by the observed MDF.

The sharpest conclusion from our modeling is that the observed MDF implies a rapidly declining SFH, with $\tauSFH \sim 0.4$ Gyr, in addition to an eventual truncation.  This conclusion is driven by the turnover and slow decline of the MDF, as opposed to the high peak and rapid cutoff predicted for models with roughly constant SFR (\citetWAF{}). As discussed in Section \ref{sec:alt_model_fits}, clearly ruling out this model would require more precise \FeHtext\ values for the highest metallicity stars in \EriII, but the predicted MDF shape for a constant SFR clearly differs from that inferred in most studies of low-luminosity dwarfs. Returning to Figure \ref{fig:Fiducial_Model_Fit}, we note that the portion of the MDF beyond the maximum at $\FeH \sim -2.25$ is populated by by stars whose posterior $p(\FeH)$ peaks at higher \FeHtext, not by the long tails of stars whose most probably \FeHtext\ is lower. Thus, there is no indication in the data that the smooth turnover of the MDF (as opposed to a sharp cutoff) is caused by observational scatter.

There are two ways in which departures from a one-zone model could explain a turnover in the MDF without a rapidly declining SFH or otherwise bias our results: (1) spatial metallicity gradients (2) stochastic enrichment events. We discuss these scenarios below, but note that either of these would require additional degrees of freedom to model. If it remains consistent with future data, the parsimony of the 4-parameter one-zone model is an argument in its favor.

\subsubsection{Spatial Gradients}
Many, though not all, dwarf galaxies are known to host mild radial stellar metallicity gradients of $\nabla_\FeH\sim-0.1$ dex/$r_h$ \citep[see][and references therein]{taibi:2022}. The presence of a metallicity gradient in \EriII\ could impact our results in one of two ways. Our sample may be biased to higher metallicity because our CaHK measurements only include stars within $\sim$1 $r_h$ thereby missing the most metal-poor stars at large radii. Alternatively, the shape of \EriII's MDF may be altered by the inclusion of stars at a range of radii that do not share identical chemical enrichment histories, thus violating the assumption of a one-zone chemical evolution model.

In most cases, the metallicity gradients of dwarf galaxies are thought to be primarily generated by feedback-driven outflows, which heat stellar orbits and preferentially drive outward migration of old stars \citep{el-badry:2016}. Because the oldest stars are also likely to be the most metal-poor, this migration can create a negative stellar metallicity gradient with more metal-rich stars at small radii and more metal-poor stars at large radii. While \citet{el-badry:2016} find feedback-driven stellar migration to be more pronounced for low-mass galaxies in a slightly larger mass regime ($M_* \sim 10^{7-9.6}$)\footnote{Specifically, \citet{el-badry:2016} find that the higher dark matter fractions and lower SFE of less massive galaxies leads to smaller fluctuations in the galactic potential and therefore weaker coupling between feedback-driven outflows and stellar kinematics.}, we cannot entirely rule out the possibility that our sample of stars in \EriII\ has been impacted by radial migration.

Alternatively, radial metallicity gradients may be indicative of radial gradients in galactic physics. For example, if the outflow mass-loading increased with radius or the SFE decreased with radius, then the central regions of the galaxy could evolve to higher \FeHtext\ than the outer regions. In principle, this could produce a small fraction of stars with \FeHtext\ beyond the peak of the MDF, creating the turnover we see in \EriII's MDF without needing to invoke an exponentially declining SFH. We have not experimented with such models, but with freedom to choose the density profile and $\eta(r)$ or $\tauSFE(r)$ we expect one could produce a range of MDF shapes. In chemical evolution studies of more massive galaxies like the Milky Way, radial variations in model parameters is indeed important and has motivated the replacement of the single one-zone model with a series of concentric one-zone models, each representing a radial annulus of the galaxy \citep[e.g.,][]{matteucci:1989, schonrich:2009, minchev:2013, johnson:2021, sharma:2021}. However, such effects are likely to be much less important for UFDs like \EriII, which formed the bulk of their stars on very short timescales and very small spatial scales.

So far, only two studies, \citet{martinez-vazquez:2021} and \citetFu{}, have attempted to measure a stellar radial metallicity gradient in \EriII. Using a sample of 67 RR Lyrae stars, \citet{martinez-vazquez:2021} measured a strong negative metallicity gradient of $-0.46$ dex/$r_h$ in the inner half-light radius of \EriII. However, a gradient of this magnitude is highly unusual for an isolated dwarf galaxy of \EriII's mass and is more characteristic of dwarf galaxies known to have experienced a past merger event \citep[e.g., Sextans, Andromeda II, Phoenix, NGC 6822;][]{taibi:2022}. Moreover, \citetFu{}, which provides the observational basis for our analysis, found no evidence for a spatial trend in stellar [Fe/H] within one half-light radius where the gradient was reported to be strongest by \citet{martinez-vazquez:2021}. The origin of this discrepancy remains uncertain and merits future investigation. Nevertheless, because no radial metallicity gradient exists within our sample, we do not believe the shape of our MDF to be substantially altered by the presence of spatial gradients in \EriII. 

It is still possible that we are biased by our centrally concentrated sample, which may not include old, metal-poor stars that formed or migrated beyond the inner half-light radius. A more spatially extended survey of stellar metallicities in \EriII\ is necessary in order to rule this possibility out. If it turns out that our current sample is missing a sizeable population of metal-poor stars, then the SFE or SFH timescale we infer may be biased high. A more spatially extended survey of stellar metallicities in \EriII\ is necessary in order to quantify the magnitude of the bias or rule this possibility out.

\subsubsection{Stochastic Supernova Enrichment}
A second possible departure from our model assumptions is Poisson sampling of the supernova population. For $M_* \sim 2\times 10^5$ M$_\odot$ the number of CCSN is $\sim$2000 and the number of SN Ia is $\sim$200. If the reservoir is fully mixed, as appears to be a reasonable assumption for ancient dwarf galaxies \citep[see][]{escala:2018}, then Poisson fluctuations would produce only minor variations in the enrichment history, at least at the high metallicity end of the MDF. However, if the galaxy is divided into smaller zones that do not efficiently share metals with each other then the number of supernovae that contribute to the composition of any given star is smaller. In this scenario, the high metallicity tail of the MDF could be populated by stars that happened to be enriched by unusually large numbers of supernovae -- most likely SN Ia because of their smaller numbers and larger Fe yield per supernova, though stochastic sampling of the IMF at late times when the SFR is low may also contribute to fluctuations in the number of CCSNe. In principle this scenario could be tested by measuring stochastic fluctuations in element ratios, following the arguments presented by \cite{griffith:2023}. While stochastic sampling may be a small effect in this study of \EriII, it is likely to be more important in lower mass UFDs where the total number of CCSN and SN Ia may be smaller and in analyses that involve additional element ratios \citep[e.g.,][]{alexander:2023}.

\subsection{Comparison to Johnson et al.\ (2022b)}
\label{sec:VICE_comparison}
Although the formulation is quite different, our method has features in common with the recently proposed method of \citet[][hereafter \citetJohnson{}]{johnson:2022b}, which also fits dwarf galaxy abundance data with one-zone chemical evolution models. The \citetJohnson{} method considers the probability that each star can be associated with each point on a model evolutionary track. Weighting these probabilities by the model SFR enforces a good match to the MDF of the data set. Our method works directly from the MDF, though the treatment of measured $P(\FeH|\text{CaHK})$ as a prior on the latent $P(\FeH')$ of each star makes the calculation resemble the likelihood calculation of \citetJohnson{}. \citetJohnson{} consider data with both \alphaFetext\ and \FeHtext\ measurements, and the \alphaFetext\ turnover provides leverage on the model timescales given the DTD of SN Ia enrichment. For \EriII\ we have been able to derive surprisingly strong constraints from $P(\FeH)$ alone, though we are aided by the turnover form of the MDF and by the known early truncation of star formation.

We suspect, but have not yet tested, that the two methods would give similar results from equivalent input data.  We use the \citetWAF{} analytic solutions while \citetJohnson{} use numerical computations from \texttt{VICE} \citep{johnson:2020}, but in principle either method could be implemented using analytic or numerical chemical evolution calculations. Our method could be generalized to model a joint $P(\alphaFe,\FeH)$ distribution, but
the \citetJohnson{} method may be simpler to implement when multiple observables per star are included.  Conversely, our approach may be better adapted to complex non-Gaussian \FeHtext\ uncertainties like those derived from CaHK photometry. Further work is merited to understand the consistency of 
these approaches and their relative strengths for different classes of observational data.

\section{Conclusion}
\label{sec:conclusion}
In this work, we use an analytic one-zone galactic chemical evolution model to fit the CaHK MDF of \EriII\ in a hierarchical Bayesian framework that appropriately accounts for non-Gaussian measurement uncertainties. Our Fiducial model achieves a good match to the observed MDF from which we infer reasonable constraints on \EriII's SFH ($\tauSFH=0.39\pm_{0.13}^{0.18}$ Gyr), SFE ($\tauSFE=27.56\pm_{12.92}^{25.14}$ Gyr), and mass-loading factor ($\eta=194.53\pm_{42.67}^{33.37}$). These results are consistent with expectations of both low SFE and high $\eta$ in low-mass galaxies and with direct estimates of \EriII's SFH from deep photometric data.

Our best-fit Fiducial model paints the following picture of \EriII's evolution. When star formation began, \EriII\ had an initial gas mass of $\sim$10$^{7}$ M$_\odot$ and continued to accrete gas vigorously but at an exponentially declining rate. Because of its low SFE and the presence of strong stellar feedback which drives ISM gas out of its shallow potential well, only a small fraction of the accreted gas is converted into stars. The production of Fe is dominated at early times by CCSNe and at late times by SN Ia, though feedback-induced galactic winds remove $>$90\% of all Fe from the galaxy, resulting in \EriII's low final metallicity. Gas loss from these large outflows outpaces gas accretion, resulting in an exponentially declining SFH that truncates at $\sim$1.4 Gyr -- likely as a result of reionization evaporating its remaining gas supply.

In addition to our Fiducial model, we consider several alternative models to build physical intuition and test specific formation scenarios (e.g., a constant SFR). These models, by-and-large, yield less natural fits to Eri II's MDF compared to the Fiducial model, though in some cases they remain statistically acceptable because of the uncertainties of the stellar \FeHtext\ measurements. Additional investigation is required to evaluate the role that spatial variation and stochastic SN enrichment may play in \EriII's MDF. Similarly, given its low mass and early star formation, a more physically-motivated treatment of reionization is warranted. It is encouraging that the stellar MDF alone gives informative constraints on the evolution of \EriII\ within the framework of the Fiducial model.

Regarding future observations of \EriII\ and other UFDs, we stress the importance of acquiring precise spectroscopic abundances of not just the lowest metallicity stars, but also -- and especially -- stars at the high-metallicity end of the MDF. The metallicities and element abundance ratios of these stars will provide some of the strongest constraints on the inferred evolution of their host galaxies.

\section*{Acknowledgements}
We thank Ellie Abrahams, Ani Chiti, Alex Ji, James Johnson, Andrey Kravtsov, and Alessandro Savino for insightful discussion on this work. 
NRS is grateful for the hospitality of the Ohio State University Department of Astronomy during the conclusion of this work. NRS and SFW acknowledge support from the NSF GRFP under grants DGE 1752814 and DGE 2146752. SWF also acknowledges support from the Paul \& Daisy Soros Fellowship.
NRS, DRW, and SWF acknowledge support from HST-GO-15901, HST-AR-16159, HST-GO-16226 and JWST DD-ERS-1334 from the Space Telescope Science Institute, which is operated by AURA, Inc., under NASA contract NAS5-26555. 
DHW acknowledges the support and hospitality of the Miller Institute for Basic Research during the initiation of this work and the support of NSF grant AST-19009841. 
The computations in this paper were partially run on the Savio computational cluster resource provided by the Berkeley Research Computing Program at the University of California, Berkeley. \\ \\

\noindent\textit{Software:} 
\texttt{astropy} \citep{astropycollaboration:2013, astropycollaboration:2018, astropycollaboration:2022},
\texttt{corner} \citep{foreman-mackey:2016},
\texttt{matplotlib} \citep{Hunter:2007},
\texttt{numpy} \citep{walt:2011, harris:2020},
\texttt{pandas} \citep{mckinney:2010, reback:2022},
\texttt{pocoMC} \citep{karamanis:2022, pocomc:2022},
\texttt{scipy} \citep{virtanen:2020},

\section*{Data Availability}
The data and code underlying this article are available in a Github repository, at \url{https://github.com/NathanSandford/ChemWAF}.



\bibliographystyle{mnras}
\bibliography{Eri_II_MDF} % if your bibtex file is called example.bib

\begin{thebibliography}{}
\makeatletter
\relax
\def\mn@urlcharsother{\let\do\@makeother \do\$\do\&\do\#\do\^\do\_\do\%\do\~}
\def\mn@doi{\begingroup\mn@urlcharsother \@ifnextchar [ {\mn@doi@}
  {\mn@doi@[]}}
\def\mn@doi@[#1]#2{\def\@tempa{#1}\ifx\@tempa\@empty \href
  {http://dx.doi.org/#2} {doi:#2}\else \href {http://dx.doi.org/#2} {#1}\fi
  \endgroup}
\def\mn@eprint#1#2{\mn@eprint@#1:#2::\@nil}
\def\mn@eprint@arXiv#1{\href {http://arxiv.org/abs/#1} {{\tt arXiv:#1}}}
\def\mn@eprint@dblp#1{\href {http://dblp.uni-trier.de/rec/bibtex/#1.xml}
  {dblp:#1}}
\def\mn@eprint@#1:#2:#3:#4\@nil{\def\@tempa {#1}\def\@tempb {#2}\def\@tempc
  {#3}\ifx \@tempc \@empty \let \@tempc \@tempb \let \@tempb \@tempa \fi \ifx
  \@tempb \@empty \def\@tempb {arXiv}\fi \@ifundefined
  {mn@eprint@\@tempb}{\@tempb:\@tempc}{\expandafter \expandafter \csname
  mn@eprint@\@tempb\endcsname \expandafter{\@tempc}}}

\bibitem[\protect\citeauthoryear{Aghanim et~al.,}{Aghanim
  et~al.}{2020}]{planck:2020}
Aghanim N.,  et~al., 2020, \mn@doi [Astronomy and Astrophysics]
  {10.1051/0004-6361/201833910}, 641, A6

\bibitem[\protect\citeauthoryear{Alexander, Vincenzo, Ji, Richstein, Jordan  \&
  Gibson}{Alexander et~al.}{2023}]{alexander:2023}
Alexander R.~K.,  Vincenzo F.,  Ji A.~P.,  Richstein H.,  Jordan C.~J.,
  Gibson B.~K.,  2023, \mn@doi [Monthly Notices of the Royal Astronomical
  Society] {10.1093/mnras/stad1312}, 522, 5415

\bibitem[\protect\citeauthoryear{Andrews, Weinberg, Sch{\"o}nrich  \&
  Johnson}{Andrews et~al.}{2017}]{andrews:2017}
Andrews B.~H.,  Weinberg D.~H.,  Sch{\"o}nrich R.,   Johnson J.~A.,  2017,
  \mn@doi [The Astrophysical Journal] {10.3847/1538-4357/835/2/224}, 835, 224

\bibitem[\protect\citeauthoryear{Asplund, Grevesse, Sauval  \& Scott}{Asplund
  et~al.}{2009}]{asplund:2009}
Asplund M.,  Grevesse N.,  Sauval A.~J.,   Scott P.,  2009, \mn@doi [Annual
  Review of Astronomy and Astrophysics]
  {10.1146/annurev.astro.46.060407.145222}, 47, 481

\bibitem[\protect\citeauthoryear{{Astropy Collaboration} et~al.,}{{Astropy
  Collaboration} et~al.}{2013}]{astropycollaboration:2013}
{Astropy Collaboration} et~al., 2013, \mn@doi [Astronomy and Astrophysics]
  {10.1051/0004-6361/201322068}, 558, A33

\bibitem[\protect\citeauthoryear{{Astropy Collaboration} et~al.,}{{Astropy
  Collaboration} et~al.}{2018}]{astropycollaboration:2018}
{Astropy Collaboration} et~al., 2018, \mn@doi [The Astronomical Journal]
  {10.3847/1538-3881/aabc4f}, 156, 123

\bibitem[\protect\citeauthoryear{{Astropy Collaboration} et~al.,}{{Astropy
  Collaboration} et~al.}{2022}]{astropycollaboration:2022}
{Astropy Collaboration} et~al., 2022, \mn@doi [The Astrophysical Journal]
  {10.3847/1538-4357/ac7c74}, 935, 167

\bibitem[\protect\citeauthoryear{Battaglia, Taibi, Thomas  \& Fritz}{Battaglia
  et~al.}{2022}]{battaglia:2022}
Battaglia G.,  Taibi S.,  Thomas G.~F.,   Fritz T.~K.,  2022, \mn@doi
  [Astronomy and Astrophysics] {10.1051/0004-6361/202141528}, 657, A54

\bibitem[\protect\citeauthoryear{Bechtol et~al.,}{Bechtol
  et~al.}{2015}]{bechtol:2015}
Bechtol K.,  et~al., 2015, \mn@doi [The Astrophysical Journal]
  {10.1088/0004-637X/807/1/50}, 807, 50

\bibitem[\protect\citeauthoryear{Behroozi, Wechsler  \& Conroy}{Behroozi
  et~al.}{2013}]{behroozi:2013}
Behroozi P.~S.,  Wechsler R.~H.,   Conroy C.,  2013, \mn@doi [The Astrophysical
  Journal] {10.1088/2041-8205/762/2/L31}, 762, L31

\bibitem[\protect\citeauthoryear{Benson, Bower, Frenk, Lacey, Baugh  \&
  Cole}{Benson et~al.}{2003}]{benson:2003}
Benson A.~J.,  Bower R.~G.,  Frenk C.~S.,  Lacey C.~G.,  Baugh C.~M.,   Cole
  S.,  2003, \mn@doi [The Astrophysical Journal] {10.1086/379160}, 599, 38

\bibitem[\protect\citeauthoryear{Chevalier \& Clegg}{Chevalier \&
  Clegg}{1985}]{chevalier:1985}
Chevalier R.~A.,  Clegg A.~W.,  1985, \mn@doi [Nature] {10.1038/317044a0}, 317,
  44

\bibitem[\protect\citeauthoryear{Chisholm, Tremonti, Leitherer  \&
  Chen}{Chisholm et~al.}{2017}]{chisholm:2017}
Chisholm J.,  Tremonti C.~A.,  Leitherer C.,   Chen Y.,  2017, \mn@doi [Monthly
  Notices of the Royal Astronomical Society] {10.1093/mnras/stx1164}, 469, 4831

\bibitem[\protect\citeauthoryear{Choi, Dotter, Conroy, Cantiello, Paxton  \&
  Johnson}{Choi et~al.}{2016}]{choi:2016}
Choi J.,  Dotter A.,  Conroy C.,  Cantiello M.,  Paxton B.,   Johnson B.~D.,
  2016, \mn@doi [The Astrophysical Journal] {10.3847/0004-637X/823/2/102},
  \href {https://ui.adsabs.harvard.edu/abs/2016ApJ...823..102C} {823, 102}

\bibitem[\protect\citeauthoryear{Conroy et~al.,}{Conroy
  et~al.}{2022}]{conroy:2022}
Conroy C.,  et~al., 2022, Birth of the {{Galactic Disk Revealed}} by the {{H3
  Survey}}, \mn@doi{10.48550/arXiv.2204.02989}

\bibitem[\protect\citeauthoryear{Crnojevi{\'c}, Sand, Zaritsky, Spekkens,
  Willman  \& Hargis}{Crnojevi{\'c} et~al.}{2016}]{crnojevic:2016}
Crnojevi{\'c} D.,  Sand D.~J.,  Zaritsky D.,  Spekkens K.,  Willman B.,
  Hargis J.~R.,  2016, \mn@doi [The Astrophysical Journal]
  {10.3847/2041-8205/824/1/L14}, 824, L14

\bibitem[\protect\citeauthoryear{Dav{\'e}, Katz, Oppenheimer, Kollmeier  \&
  Weinberg}{Dav{\'e} et~al.}{2013}]{dave:2013}
Dav{\'e} R.,  Katz N.,  Oppenheimer B.~D.,  Kollmeier J.~A.,   Weinberg D.~H.,
  2013, \mn@doi [Monthly Notices of the Royal Astronomical Society]
  {10.1093/mnras/stt1274}, 434, 2645

\bibitem[\protect\citeauthoryear{Dotter}{Dotter}{2016}]{dotter:2016}
Dotter A.,  2016, \mn@doi [The Astrophysical Journal Supplement Series]
  {10.3847/0067-0049/222/1/8}, 222, 8

\bibitem[\protect\citeauthoryear{Eggen, Scarlata, Skillman  \& Jaskot}{Eggen
  et~al.}{2022}]{eggen:2022}
Eggen N.~R.,  Scarlata C.,  Skillman E.,   Jaskot A.,  2022, Blow-{{Away}} in
  the {{Extreme Low-Mass Starburst Galaxy Pox}}\textasciitilde 186 (\mn@eprint
  {arxiv} {2207.02245}), \mn@doi{10.48550/arXiv.2207.02245}

\bibitem[\protect\citeauthoryear{{El-Badry}, Wetzel, Geha, Hopkins, Kere{\v s},
  Chan  \& {Faucher-Gigu{\`e}re}}{{El-Badry} et~al.}{2016}]{el-badry:2016}
{El-Badry} K.,  Wetzel A.,  Geha M.,  Hopkins P.~F.,  Kere{\v s} D.,  Chan
  T.~K.,   {Faucher-Gigu{\`e}re} C.-A.,  2016, \mn@doi [The Astrophysical
  Journal] {10.3847/0004-637X/820/2/131}, 820, 131

\bibitem[\protect\citeauthoryear{Emerick, Bryan  \& Mac~Low}{Emerick
  et~al.}{2019}]{emerick:2019}
Emerick A.,  Bryan G.~L.,   Mac~Low M.-M.,  2019, \mn@doi [Monthly Notices of
  the Royal Astronomical Society] {10.1093/mnras/sty2689}, 482, 1304

\bibitem[\protect\citeauthoryear{Escala et~al.,}{Escala
  et~al.}{2018}]{escala:2018}
Escala I.,  et~al., 2018, \mn@doi [Monthly Notices of the Royal Astronomical
  Society] {10.1093/mnras/stx2858}, 474, 2194

\bibitem[\protect\citeauthoryear{Finlator \& Dav{\'e}}{Finlator \&
  Dav{\'e}}{2008}]{finlator:2008}
Finlator K.,  Dav{\'e} R.,  2008, \mn@doi [Monthly Notices of the Royal
  Astronomical Society] {10.1111/j.1365-2966.2008.12991.x}, 385, 2181

\bibitem[\protect\citeauthoryear{{Foreman-Mackey}}{{Foreman-Mackey}}{2016}]{foreman-mackey:2016}
{Foreman-Mackey} D.,  2016, \mn@doi [Journal of Open Source Software]
  {10.21105/joss.00024}, 1, 24

\bibitem[\protect\citeauthoryear{Fu et~al.,}{Fu et~al.}{2022}]{fu:2022}
Fu S.~W.,  et~al., 2022, \mn@doi [The Astrophysical Journal]
  {10.3847/1538-4357/ac3665}, 925, 6

\bibitem[\protect\citeauthoryear{Gallart et~al.,}{Gallart
  et~al.}{2021}]{gallart:2021}
Gallart C.,  et~al., 2021, \mn@doi [The Astrophysical Journal]
  {10.3847/1538-4357/abddbe}, 909, 192

\bibitem[\protect\citeauthoryear{Griffith, Johnson, Weinberg, Ilyin, Johnson,
  {Rodriguez-Martinez}  \& Strassmeier}{Griffith et~al.}{2023}]{griffith:2023}
Griffith E.~J.,  Johnson J.~A.,  Weinberg D.~H.,  Ilyin I.,  Johnson J.~W.,
  {Rodriguez-Martinez} R.,   Strassmeier K.~G.,  2023, \mn@doi [The
  Astrophysical Journal] {10.3847/1538-4357/aca659}, 944, 47

\bibitem[\protect\citeauthoryear{Harris et~al.,}{Harris
  et~al.}{2020}]{harris:2020}
Harris C.~R.,  et~al., 2020, \mn@doi [Nature] {10.1038/s41586-020-2649-2}, 585,
  357

\bibitem[\protect\citeauthoryear{Hopkins, Quataert  \& Murray}{Hopkins
  et~al.}{2012}]{hopkins:2012}
Hopkins P.~F.,  Quataert E.,   Murray N.,  2012, \mn@doi [Monthly Notices of
  the Royal Astronomical Society] {10.1111/j.1365-2966.2012.20593.x}, 421, 3522

\bibitem[\protect\citeauthoryear{Hopkins, Kere{\v s}, O{\~n}orbe,
  {Faucher-Gigu{\`e}re}, Quataert, Murray  \& Bullock}{Hopkins
  et~al.}{2014}]{hopkins:2014}
Hopkins P.~F.,  Kere{\v s} D.,  O{\~n}orbe J.,  {Faucher-Gigu{\`e}re} C.-A.,
  Quataert E.,  Murray N.,   Bullock J.~S.,  2014, \mn@doi [Monthly Notices of
  the Royal Astronomical Society] {10.1093/mnras/stu1738}, 445, 581

\bibitem[\protect\citeauthoryear{Hopkins et~al.,}{Hopkins
  et~al.}{2018}]{hopkins:2018}
Hopkins P.~F.,  et~al., 2018, \mn@doi [Monthly Notices of the Royal
  Astronomical Society] {10.1093/mnras/sty1690}, 480, 800

\bibitem[\protect\citeauthoryear{Hunter}{Hunter}{2007}]{Hunter:2007}
Hunter J.~D.,  2007, \mn@doi [Computing in Science \& Engineering]
  {10.1109/MCSE.2007.55}, 9, 90

\bibitem[\protect\citeauthoryear{Johnson \& Weinberg}{Johnson \&
  Weinberg}{2020}]{johnson:2020}
Johnson J.~W.,  Weinberg D.~H.,  2020, \mn@doi [Monthly Notices of the Royal
  Astronomical Society] {10.1093/mnras/staa2431}, 498, 1364

\bibitem[\protect\citeauthoryear{Johnson et~al.,}{Johnson
  et~al.}{2021}]{johnson:2021}
Johnson J.~W.,  et~al., 2021, \mn@doi [Monthly Notices of the Royal
  Astronomical Society] {10.1093/mnras/stab2718}, 508, 4484

\bibitem[\protect\citeauthoryear{Johnson, Kochanek  \& Stanek}{Johnson
  et~al.}{2022a}]{johnson:2022a}
Johnson J.~W.,  Kochanek C.~S.,   Stanek K.~Z.,  2022a, Binaries Drive High
  {{Type Ia}} Supernova Rates in Dwarf Galaxies,
  \mn@doi{10.48550/arXiv.2210.01818}

\bibitem[\protect\citeauthoryear{Johnson et~al.,}{Johnson
  et~al.}{2022b}]{johnson:2022b}
Johnson J.~W.,  et~al., 2022b, Dwarf Galaxy Archaeology from Chemical
  Abundances and Star Formation Histories, \mn@doi{10.48550/arXiv.2210.01816}

\bibitem[\protect\citeauthoryear{Karamanis, Nabergoj, Beutler, Peacock  \&
  Seljak}{Karamanis et~al.}{2022a}]{pocomc:2022}
Karamanis M.,  Nabergoj D.,  Beutler F.,  Peacock J.,   Seljak U.,  2022a,
  \mn@doi [The Journal of Open Source Software] {10.21105/joss.04634}, 7, 4634

\bibitem[\protect\citeauthoryear{Karamanis, Beutler, Peacock, Nabergoj  \&
  Seljak}{Karamanis et~al.}{2022b}]{karamanis:2022}
Karamanis M.,  Beutler F.,  Peacock J.~A.,  Nabergoj D.,   Seljak U.,  2022b,
  \mn@doi [Monthly Notices of the Royal Astronomical Society]
  {10.1093/mnras/stac2272}, 516, 1644

\bibitem[\protect\citeauthoryear{Kirby, Lanfranchi, Simon, Cohen  \&
  Guhathakurta}{Kirby et~al.}{2011}]{kirby:2011a}
Kirby E.~N.,  Lanfranchi G.~A.,  Simon J.~D.,  Cohen J.~G.,   Guhathakurta P.,
  2011, \mn@doi [The Astrophysical Journal] {10.1088/0004-637X/727/2/78}, 727,
  78

\bibitem[\protect\citeauthoryear{Koposov, Belokurov, Torrealba  \&
  Evans}{Koposov et~al.}{2015}]{koposov:2015}
Koposov S.~E.,  Belokurov V.,  Torrealba G.,   Evans N.~W.,  2015, \mn@doi [The
  Astrophysical Journal] {10.1088/0004-637X/805/2/130}, 805, 130

\bibitem[\protect\citeauthoryear{Kroupa}{Kroupa}{2001}]{kroupa:2001}
Kroupa P.,  2001, \mn@doi [Monthly Notices of the Royal Astronomical Society]
  {10.1046/j.1365-8711.2001.04022.x}, 322, 231

\bibitem[\protect\citeauthoryear{Lacchin, Matteucci, Vincenzo  \&
  Palla}{Lacchin et~al.}{2020}]{lacchin:2020}
Lacchin E.,  Matteucci F.,  Vincenzo F.,   Palla M.,  2020, \mn@doi [Monthly
  Notices of the Royal Astronomical Society] {10.1093/mnras/staa585}, 495, 3276

\bibitem[\protect\citeauthoryear{Lanfranchi \& Matteucci}{Lanfranchi \&
  Matteucci}{2004}]{lanfranchi:2004}
Lanfranchi G.~A.,  Matteucci F.,  2004, \mn@doi [Monthly Notices of the Royal
  Astronomical Society] {10.1111/j.1365-2966.2004.07877.x}, 351, 1338

\bibitem[\protect\citeauthoryear{Lanfranchi \& Matteucci}{Lanfranchi \&
  Matteucci}{2007}]{lanfranchi:2007}
Lanfranchi G.~A.,  Matteucci F.,  2007, \mn@doi [Astronomy and Astrophysics]
  {10.1051/0004-6361:20066576}, 468, 927

\bibitem[\protect\citeauthoryear{Lanfranchi \& Matteucci}{Lanfranchi \&
  Matteucci}{2010}]{lanfranchi:2010}
Lanfranchi G.~A.,  Matteucci F.,  2010, \mn@doi [Astronomy and Astrophysics]
  {10.1051/0004-6361/200913045}, 512, A85

\bibitem[\protect\citeauthoryear{Lanfranchi, Matteucci  \& Cescutti}{Lanfranchi
  et~al.}{2006}]{lanfranchi:2006}
Lanfranchi G.~A.,  Matteucci F.,   Cescutti G.,  2006, \mn@doi [Astronomy and
  Astrophysics] {10.1051/0004-6361:20054627}, 453, 67

\bibitem[\protect\citeauthoryear{Leroy, Walter, Brinks, Bigiel, {de Blok},
  Madore  \& Thornley}{Leroy et~al.}{2008}]{leroy:2008}
Leroy A.~K.,  Walter F.,  Brinks E.,  Bigiel F.,  {de Blok} W. J.~G.,  Madore
  B.,   Thornley M.~D.,  2008, \mn@doi [The Astronomical Journal]
  {10.1088/0004-6256/136/6/2782}, 136, 2782

\bibitem[\protect\citeauthoryear{Li et~al.,}{Li et~al.}{2017}]{li:2017}
Li T.~S.,  et~al., 2017, \mn@doi [The Astrophysical Journal]
  {10.3847/1538-4357/aa6113}, 838, 8

\bibitem[\protect\citeauthoryear{Maoz \& Graur}{Maoz \&
  Graur}{2017}]{maoz:2017}
Maoz D.,  Graur O.,  2017, \mn@doi [The Astrophysical Journal]
  {10.3847/1538-4357/aa8b6e}, 848, 25

\bibitem[\protect\citeauthoryear{Maoz, Mannucci  \& Brandt}{Maoz
  et~al.}{2012}]{maoz:2012}
Maoz D.,  Mannucci F.,   Brandt T.~D.,  2012, \mn@doi [Monthly Notices of the
  Royal Astronomical Society] {10.1111/j.1365-2966.2012.21871.x}, 426, 3282

\bibitem[\protect\citeauthoryear{{Mart{\'i}nez-V{\'a}zquez}
  et~al.,}{{Mart{\'i}nez-V{\'a}zquez} et~al.}{2021}]{martinez-vazquez:2021}
{Mart{\'i}nez-V{\'a}zquez} C.~E.,  et~al., 2021, \mn@doi [Monthly Notices of
  the Royal Astronomical Society] {10.1093/mnras/stab2493}, 508, 1064

\bibitem[\protect\citeauthoryear{Matteucci \& Francois}{Matteucci \&
  Francois}{1989}]{matteucci:1989}
Matteucci F.,  Francois P.,  1989, \mn@doi [Monthly Notices of the Royal
  Astronomical Society] {10.1093/mnras/239.3.885}, 239, 885

\bibitem[\protect\citeauthoryear{McKinney}{McKinney}{2010}]{mckinney:2010}
McKinney W.,  2010, in {van der Walt} S.,  Millman J.,  eds, Proceedings of the
  9th {{Python}} in {{Science Conference}}. pp 56--61,
  \mn@doi{10.25080/Majora-92bf1922-00a}

\bibitem[\protect\citeauthoryear{McQuinn, {van Zee}  \& Skillman}{McQuinn
  et~al.}{2019}]{mcquinn:2019}
McQuinn {\relax Kristen}. B.~W.,  {van Zee} L.,   Skillman E.~D.,  2019,
  \mn@doi [The Astrophysical Journal] {10.3847/1538-4357/ab4c37}, 886, 74

\bibitem[\protect\citeauthoryear{Minchev, Chiappini  \& Martig}{Minchev
  et~al.}{2013}]{minchev:2013}
Minchev I.,  Chiappini C.,   Martig M.,  2013, Astronomy and Astrophysics, 558,
  A9

\bibitem[\protect\citeauthoryear{Mitchell, Schaye, Bower  \& Crain}{Mitchell
  et~al.}{2020}]{mitchell:2020}
Mitchell P.~D.,  Schaye J.,  Bower R.~G.,   Crain R.~A.,  2020, \mn@doi
  [Monthly Notices of the Royal Astronomical Society] {10.1093/mnras/staa938},
  494, 3971

\bibitem[\protect\citeauthoryear{Muratov, Kere{\v s}, {Faucher-Gigu{\`e}re},
  Hopkins, Quataert  \& Murray}{Muratov et~al.}{2015}]{muratov:2015}
Muratov A.~L.,  Kere{\v s} D.,  {Faucher-Gigu{\`e}re} C.-A.,  Hopkins P.~F.,
  Quataert E.,   Murray N.,  2015, \mn@doi [Monthly Notices of the Royal
  Astronomical Society] {10.1093/mnras/stv2126}, 454, 2691

\bibitem[\protect\citeauthoryear{Murray, Quataert  \& Thompson}{Murray
  et~al.}{2005}]{murray:2005}
Murray N.,  Quataert E.,   Thompson T.~A.,  2005, \mn@doi [The Astrophysical
  Journal] {10.1086/426067}, 618, 569

\bibitem[\protect\citeauthoryear{Murray, Quataert  \& Thompson}{Murray
  et~al.}{2010}]{murray:2010}
Murray N.,  Quataert E.,   Thompson T.~A.,  2010, \mn@doi [The Astrophysical
  Journal] {10.1088/0004-637X/709/1/191}, 709, 191

\bibitem[\protect\citeauthoryear{Pandya et~al.,}{Pandya
  et~al.}{2021}]{pandya:2021}
Pandya V.,  et~al., 2021, \mn@doi [Monthly Notices of the Royal Astronomical
  Society] {10.1093/mnras/stab2714}, 508, 2979

\bibitem[\protect\citeauthoryear{Peeples \& Shankar}{Peeples \&
  Shankar}{2011}]{peeples:2011}
Peeples M.~S.,  Shankar F.,  2011, \mn@doi [Monthly Notices of the Royal
  Astronomical Society] {10.1111/j.1365-2966.2011.19456.x}, 417, 2962

\bibitem[\protect\citeauthoryear{Reback et~al.,}{Reback
  et~al.}{2022}]{reback:2022}
Reback J.,  et~al., 2022, Pandas-Dev/Pandas: {{Pandas}} 1.4.3, Zenodo,
  \mn@doi{10.5281/zenodo.6702671}

\bibitem[\protect\citeauthoryear{Rodr{\'i}guez, Maoz  \& Nakar}{Rodr{\'i}guez
  et~al.}{2022}]{rodriguez:2022}
Rodr{\'i}guez {\'O}.,  Maoz D.,   Nakar E.,  2022, The {{Iron Yield}} of
  {{Core-collapse Supernovae}}, \mn@doi{10.48550/arXiv.2209.05552}

\bibitem[\protect\citeauthoryear{Romano, Bellazzini, Starkenburg  \&
  Leaman}{Romano et~al.}{2015}]{romano:2015}
Romano D.,  Bellazzini M.,  Starkenburg E.,   Leaman R.,  2015, \mn@doi
  [Monthly Notices of the Royal Astronomical Society] {10.1093/mnras/stu2427},
  446, 4220

\bibitem[\protect\citeauthoryear{Sandford, Weisz  \& Ting}{Sandford
  et~al.}{2020}]{sandford:2020b}
Sandford N.~R.,  Weisz D.~R.,   Ting Y.-S.,  2020, \mn@doi [The Astrophysical
  Journal Supplement Series] {10.3847/1538-4365/ab9cb0}, 249, 24

\bibitem[\protect\citeauthoryear{Sch{\"o}nrich \& Binney}{Sch{\"o}nrich \&
  Binney}{2009}]{schonrich:2009}
Sch{\"o}nrich R.,  Binney J.,  2009, \mn@doi [Monthly Notices of the Royal
  Astronomical Society] {10.1111/j.1365-2966.2009.14750.x}, 396, 203

\bibitem[\protect\citeauthoryear{Sharma, Hayden  \& {Bland-Hawthorn}}{Sharma
  et~al.}{2021}]{sharma:2021}
Sharma S.,  Hayden M.~R.,   {Bland-Hawthorn} J.,  2021, \mn@doi [Monthly
  Notices of the Royal Astronomical Society] {10.1093/mnras/stab2015}, 507,
  5882

\bibitem[\protect\citeauthoryear{Simon}{Simon}{2019}]{simon:2019}
Simon J.~D.,  2019, \mn@doi [Annual Review of Astronomy and Astrophysics]
  {10.1146/annurev-astro-091918-104453}, 57, 375

\bibitem[\protect\citeauthoryear{Simon et~al.,}{Simon
  et~al.}{2021}]{simon:2021}
Simon J.~D.,  et~al., 2021, \mn@doi [The Astrophysical Journal]
  {10.3847/1538-4357/abd31b}, 908, 18

\bibitem[\protect\citeauthoryear{Somerville \& Dav{\'e}}{Somerville \&
  Dav{\'e}}{2015}]{somerville:2015}
Somerville R.~S.,  Dav{\'e} R.,  2015, \mn@doi [Annual Review of Astronomy and
  Astrophysics] {10.1146/annurev-astro-082812-140951}, 53, 51

\bibitem[\protect\citeauthoryear{Taibi, Battaglia, Leaman, Brooks, Riggs,
  Munshi, Revaz  \& Jablonka}{Taibi et~al.}{2022}]{taibi:2022}
Taibi S.,  Battaglia G.,  Leaman R.,  Brooks A.,  Riggs C.,  Munshi F.,  Revaz
  Y.,   Jablonka P.,  2022, \mn@doi [Astronomy and Astrophysics]
  {10.1051/0004-6361/202243508}, 665, A92

\bibitem[\protect\citeauthoryear{Vincenzo, Matteucci, Vattakunnel  \&
  Lanfranchi}{Vincenzo et~al.}{2014}]{vincenzo:2014}
Vincenzo F.,  Matteucci F.,  Vattakunnel S.,   Lanfranchi G.~A.,  2014, \mn@doi
  [Monthly Notices of the Royal Astronomical Society] {10.1093/mnras/stu710},
  441, 2815

\bibitem[\protect\citeauthoryear{Virtanen et~al.,}{Virtanen
  et~al.}{2020}]{virtanen:2020}
Virtanen P.,  et~al., 2020, \mn@doi [Nature Methods]
  {10.1038/s41592-019-0686-2}, 17, 261

\bibitem[\protect\citeauthoryear{Weinberg, Andrews  \& Freudenburg}{Weinberg
  et~al.}{2017}]{weinberg:2017}
Weinberg D.~H.,  Andrews B.~H.,   Freudenburg J.,  2017, \mn@doi [The
  Astrophysical Journal] {10.3847/1538-4357/837/2/183}, 837, 183

\bibitem[\protect\citeauthoryear{Zoutendijk et~al.,}{Zoutendijk
  et~al.}{2020}]{zoutendijk:2020}
Zoutendijk S.~L.,  et~al., 2020, \mn@doi [A\&A] {10.1051/0004-6361/201936155},
  635, A107

\bibitem[\protect\citeauthoryear{Zoutendijk, Brinchmann, Bouch{\'e}, den Brok,
  Krajnovi{\'c}, Kuijken, Maseda  \& Schaye}{Zoutendijk
  et~al.}{2021}]{zoutendijk:2021}
Zoutendijk S.~L.,  Brinchmann J.,  Bouch{\'e} N.~F.,  den Brok M.,
  Krajnovi{\'c} D.,  Kuijken K.,  Maseda M.~V.,   Schaye J.,  2021, \mn@doi
  [A\&A] {10.1051/0004-6361/202040239}, 651, A80

\bibitem[\protect\citeauthoryear{van~der Walt, Colbert  \& Varoquaux}{van~der
  Walt et~al.}{2011}]{walt:2011}
van~der Walt S.,  Colbert S.~C.,   Varoquaux G.,  2011, \mn@doi [Computing in
  Science \& Engineering] {10.1109/MCSE.2011.37}, 13, 22

\makeatother
\end{thebibliography}

%\begin{thebibliography}{99}
%\end{thebibliography}

%%%%%%%%%%%%%%%%%%%%%%%%%%%%%%%%%%%%%%%%%%%%%%%%%%

%%%%%%%%%%%%%%%%% APPENDICES %%%%%%%%%%%%%%%%%%%%%

\appendix

\section{Isolating Parameter Influences on the Model MDF}
\label{app:parameter_dependence}
\begin{figure*} 
    \begin{center}
    \includegraphics[width=\textwidth]{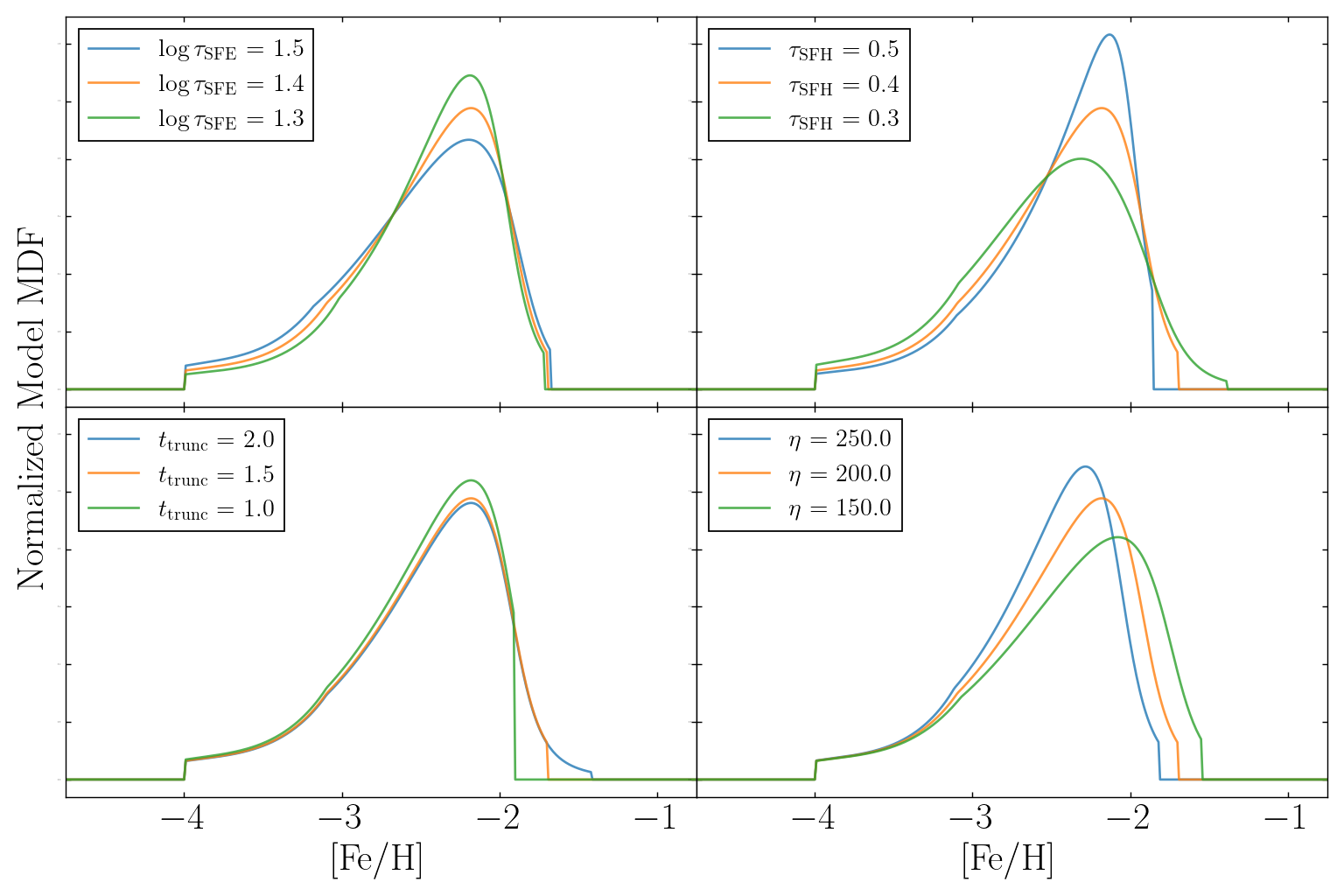}
    \caption{
        Predicted MDF of the Fiducial model as each of the free model parameters is individually increased (blue) and decreased (green) from the approximate best-fit value (orange).
        \label{fig:MDF_Parameter_Dependence1}
    }
    \end{center}
\end{figure*}

The ability of \EriII's MDF to constrain the model parameters can be understood by investigating how each parameter changes the predicted model MDF. In Figure \ref{fig:MDF_Parameter_Dependence1}, we show how the MDF changes as we increase and decrease each parameter from the best-fit Fiducial model. 
With all other parameters held fixed, increasing \logtauSFEtext\ results in a higher CC equilibrium abundance $\FeH_\text{eq}^\text{cc}$ and a longer timescale CC equilibrium timescale $\tau_\text{Fe,eq}^\text{cc}$ (Equations \ref{eq:FeHCCEq}--\ref{eq:tauDep}). This leads to a broader MDF with a more extended low-metallicity tail (top-left panel). In the SFE regime of \EriII, changing \logtauSFEtext\ does not strongly change the location of the MDFs peak.
Increasing \tauSFHtext\ has roughly the opposite effect of \logtauSFEtext, decreasing both $\FeH_\text{eq}^\text{cc}$ and $\tau_\text{Fe,eq}^\text{cc}$ and resulting in a narrower MDF with a smaller low-metallicity tail (top-right panel). While the similarities in impact between \tauSFHtext\ and \tauSFEtext\ lead to the covariance in their posteriors seen in Figure \ref{fig:Fiducial_Corner}, they are not fully degenerate. Unlike for \tauSFEtext, decreasing \tauSFHtext\ shifts the peak of the MDF to lower metallicity. Furthermore, a more extended SFH leads to an MDF that is more sharply truncated at the high-metallicity end by the abrupt end to star formation at \ttrunctext.
In comparison, the effect of $\eta$ on the shape of the MDF is more distinct (bottom-right panel). Increasing the mass-loading factor removes more metals from the galaxy, slowing the rate of enrichment and decreasing the final metallicity that the system evolves to. Higher outflows result in a narrower MDF with a lower metallicity peak and maximum \FeHtext. Changing $\eta$, however, has little impact on the low-metallicity tail of the MDF ($\FeH<-3.5$).
The direct impact of \ttrunctext\ itself is more subtle than the aforementioned parameters, because it is only responsible for the truncation of the MDF at higher metallicities (and the induced re-normalization; bottom-left panel). In our models of \EriII, \tauSFHtext\ is sufficiently short that the SFR is quite low at the time of truncation, and thus the portion of the MDF that is truncated is small. It is therefore understandable that \ttrunctext\ is prior dominated in our fits; there is no clear signature in \EriII's MDF indicating the abrupt cessation of star formation.

\begin{figure*} 
    \begin{center}
    \includegraphics[width=\textwidth]{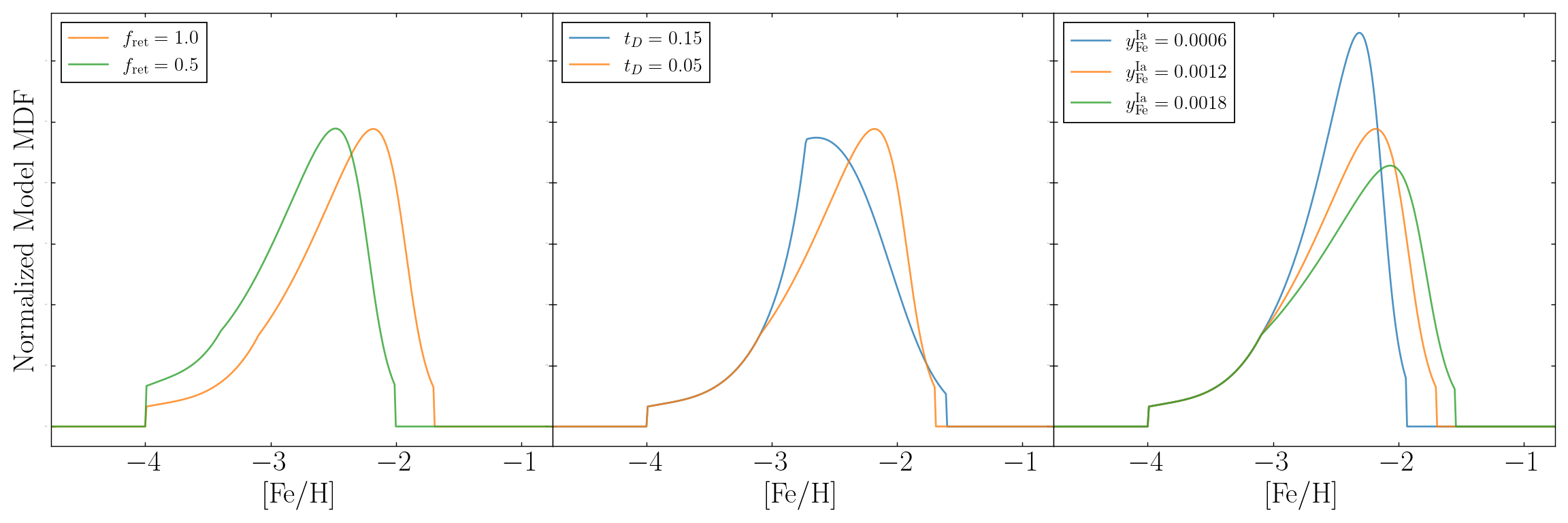}
    \caption{
        Predicted MDF of the best-fit Fiducial model (orange) compared to the predicted MDF when different values of $f_\text{ret}$, $t_D$, and $y_{\text{Fe}}^{\text{Ia}}$\ are adopted. 
        \label{fig:MDF_Parameter_Dependence2}
    }
    \end{center}
\end{figure*}

In Figure \ref{fig:MDF_Parameter_Dependence2}, we illustrate how varying a handful of model parameters that were held fixed in the Fiducial model, including \frettext, $t_D$, and \yFeIatext, influences the predicted MDF in Figure \ref{fig:MDF_Parameter_Dependence2}.
Reducing \frettext\ effectively reduces the yield of all SNe, which shifts the entire MDF to lower metallicities (left panel). A factor of two reduction as shown here results in a 0.3 dex shift to lower metallicity.
When $t_D$ is increased, SN Ia contribute less to the enrichment of the system overall and especially at early times, resulting in an MDF with a lower-metallicity peak (middle panel). The shape of the MDF below $\FeH<-2.5$ when $t_D=0.15$ Gyr is due to the model approaching the CC equilibrium metallicity, which it does on a timescale roughly equivalent to the minimum SN Ia time delay ($\tau_\text{Fe,eq}^\text{cc}\sim t_D$).
Increasing \yFeIatext\ has much the same effect as decreasing $\eta$. Large SN Ia yields drives the system to higher metallicity for $t>t_D$, resulting in broader, more metal-rich MDFs. This degeneracy with $\eta$ explains why the Enhanced SN Ia model required such large mass-loading factors to reproduce the \EriII\ MDF.

%%%%%%%%%%%%%%%%%%%%%%%%%%%%%%%%%%%%%%%%%%%%%%%%%%

% Don't change these lines
\bsp	% typesetting comment
\label{lastpage}
\end{document}